\documentclass[12pt]{article}

\usepackage{geometry}
\usepackage{microtype}
\usepackage{amssymb}
\usepackage{amsmath}
\usepackage[small]{titlesec}
\usepackage{bm}
\usepackage{mathrsfs}
\usepackage{fancyhdr}
\usepackage{fourier-orns}
\usepackage[mathcal]{euscript}
\usepackage{indentfirst}
\usepackage{forloop}
\usepackage{etoolbox}
\usepackage[symbol=\!]{footnotebackref} 

\makeatletter
\renewcommand*{\@fnsymbol}[1]{\ensuremath{\ifcase#1\or \text{\starredbullet} \else\@ctrerr\fi}}
\makeatother
\renewcommand{\d}{{\mathrm d}}
\newcommand{\im}{\operatorname{im}}
\newcommand{\D}{\mathcal{D}}

\newcommand{\A}{\mathcal{A}}
\newcommand{\F}{\mathcal{F}}
\newcommand{\W}{\mathcal{W}}
\newcommand{\Q}{\mathcal{Q}}

\renewcommand{\L}{\mathscr{L}}
\newcommand{\R}{\mathbf{R}}

\newcommand{\ad}{{\dot{\alpha}}}
\newcommand{\bd}{{\dot{\beta}}}
\newcommand{\gd}{{\dot{\gamma}}}
\newcommand{\md}{{\dot{\mu}}}
\renewcommand{\sb}{{\bar{s}}}
\newcommand{\stc}{\,\overset{*}{,}\,}
\newcommand{\bdeg}{\operatorname{bdeg}}
\renewcommand{\hat}[1]{\widehat{#1}}
\renewcommand{\epsilon}{\varepsilon}
\newcommand{\janksub}[1]{{s \ldots s #1 \psi \ldots \psi}}
\newcommand{\janksubp}{\janksub{}}
\newcommand{\janksubg}{\janksub{\gamma(s, s)}}
\newcommand{\janksubf}[1]{\janksub{\sb \ldots \sb #1}}
\newcommand{\janksubfp}{\janksubf{}}
\newcommand{\janksubfs}{\janksubf{\sigma(s, \sb)}}

\newcommand{\el}{\begin{center} \decosix ~~ \decosix ~~ \decosix \end{center}}
\makeatletter
\newcommand{\ceq}{\mathrel{\rlap{\raisebox{0.3ex}{$\m@th\cdot$}}\raisebox{-0.3ex}{$\m@th\cdot$}}=}
\newcommand{\eqc}{=\mathrel{\rlap{\raisebox{0.3ex}{$\m@th\cdot$}}\raisebox{-0.3ex}{$\m@th\cdot$}}}
\makeatother
\newtoggle{eprint}
\newcommand{\inspire}[1]{[\href{https://inspirehep.net/record/#1}{\texttt{inSPIRE:#1}}]}

\newcommand{\journal}[4]{\href{#2}{#1} \textbf{#3} #4}
\newcommand{\key}{}
\newcounter{zero}
\makeatletter
\renewenvironment{thebibliography}[1]
	{\section*{\refname}%
		\@mkboth{\MakeUppercase\refname}{\MakeUppercase\refname}%
		\list{\@biblabel{\hyperlink{\key}{\@arabic\c@enumiv}}}%
		{\settowidth\labelwidth{\@biblabel{#1}}%
			\leftmargin\labelwidth
			\advance\leftmargin\labelsep
			\@openbib@code
			\usecounter{enumiv}%
			\let\p@enumiv\@empty
			\renewcommand\theenumiv{\@arabic\c@enumiv}}%
		\sloppy
		\clubpenalty4000
		\@clubpenalty \clubpenalty
		\widowpenalty4000%
		\sfcode`\.\@m}
	{\def\@noitemerr
		{\@latex@warning{Empty `thebibliography' environment}}%
		\endlist}
\makeatother
\newcommand{\targetlist}[1]{\forcsvlist\target{#1}}
\newcommand{\target}[1]{\ucite{#1}\raisebox{\ht\strutbox}{\hypertarget{#1\arabic{#1ctr}}{}}}
\LetLtxMacro{\oldcite}{\cite}
\renewcommand{\cite}[1]{\targetlist{#1}\protect\oldcite{#1}}
\newcommand{\addocc}[1]{%
	\iftoggle{eprint}{%
	\ifnumcomp{\value{#1ctr}}{>}{0}{%
		\newcounter{#1k}\setcounter{#1k}{1}%
		\addtocounter{#1ctr}{1}
		{\footnotesize (\textit{also~cited~at}~\hyperlink{#1\arabic{#1k}}{\textsf{\roman{#1k}}}%
		\forloop{#1k}{2}{\value{#1k} < \value{#1ctr}}{%
			,\;\hyperlink{#1\arabic{#1k}}{\textsf{\roman{#1k}}}%
		}).}%
	}{}}{}}
\newcounter{citecnt}
\newtoks\citetoks
\makeatletter
\newcommand{\ucite}[1]{%
	\@ifundefined{uns@cite#1}
		{\global\newcounter{#1ctr}\setcounter{#1ctr}{0}%
		\refstepcounter{citecnt}\label{citelabel@#1}%
		\expandafter\xdef\csname uns@cite#1\endcsname{\arabic{citecnt}}%
		\toks\z@=\expandafter{\the\citetoks}%
		\toks\tw@=\expandafter\expandafter\expandafter{%
		\csname uns@bibitem#1\endcsname}%
		\edef\@tempcite{\the\toks\z@\the\toks\tw@}%
		\global\citetoks=\expandafter{\@tempcite}%
		}{\refstepcounter{#1ctr}}}
\LetLtxMacro{\oldbibitem}{\bibitem}
\renewcommand{\bibitem}[2]{%
	\@namedef{uns@bibitem#1}{\renewcommand{\key}{#1\arabic{zero}} \oldbibitem{#1} #2 \addocc{#1}}}
\makeatother

\toggletrue{eprint} 
\iftoggle{eprint}{
	\newcommand{\srrefcolor}{cyan}
	\newcommand{\srurlcolor}{cyan}
}{
	\newcommand{\srrefcolor}{black}
	\newcommand{\srurlcolor}{black}
}
\numberwithin{equation}{section}

\hypersetup{
	colorlinks = true,
	citecolor = \srrefcolor,
	linkcolor = \srrefcolor,
	urlcolor = \srurlcolor,
	bookmarksopen = true,
	bookmarksopenlevel = \maxdimen
}
\title{\textbf{Supersymmetric Tensor Hierarchies from Superspace Cohomology}}
\author{Stephen Randall\footnote{\href{mailto:srandall@berkeley.edu}{\textsf{srandall@berkeley.edu}}}}
\date{}

\begin{document}
\maketitle

\thispagestyle{fancy}
\fancyfoot[C]{}

\vspace*{-0.75cm}
\begin{center}
\textit{Department of Physics \\ University of California, Berkeley}
\end{center}
\vspace{1cm}

\abstract{In this set of lectures we give a pedagogical introduction to the way in which the nilpotency of a super-de Rham operator can be exploited for the construction of gauge theories in superspace. We begin with a discussion of how the super-geometric closure conditions can be solved by simply computing the cocycles of the super-algebra. The next couple lectures are then devoted to applying this idea to extensions of the standard super-de Rham complex. This eventually results in a geometric ``trivialization" of the consistency conditions required for non-abelian tensor hierarchies. Although this is a general conclusion, we focus specifically on the hierarchy obtained by compactifying the 3-form gauge field of 11D supergravity to 4D, $N = 1$ superspace. In the final lecture, we use the cohomological arguments developed herein to provide a geometric construction of the non-trivial Chern-Simons-type invariant in that tensor hierarchy and comment on generalizations. These lectures are based on a series of talks given at Texas A\&M University from March 21--25.}
\newpage

\thispagestyle{fancy}
\fancyfoot[C]{}

{\newgeometry{bottom=0cm,top=0cm}

\vspace*{-1.7cm}
\hypertarget{toc}{}
\renewcommand*\contentsname{{\large Lectures}}
\tableofcontents
\vspace*{\fill}}
{\clearpage
\restoregeometry}

\pagestyle{fancy}
\fancyfoot[C]{--- ~~\hyperlink{toc}{\thepage}~~ ---}
\pagenumbering{arabic}

\renewcommand*{\refname}{\vspace*{-1em}}
\section*{Introduction}
\addcontentsline{toc}{section}{Introduction}
\vspace*{-1.1em}
\setcounter{equation}{0}
\label{sec:intro}
\vspace{8pt}
\setcounter{page}{1}
\pagenumbering{roman}

When compared to bosonic cohomology, the process for building supersymmetric field-strength superfields seems quite complicated. In electromagnetism, the 2-form field-strength $F$ is formulated as the exterior derivative of the gauge 1-form $A$ so that the gauge variation suffered by $A$ (shifting by any exact 1-form) leaves $F$ invariant. In superspace we wish to accomplish the same thing, using the local contractibility of the space to build gauge forms from the closure of their field-strengths. Although the idea is clear, in practice this is complicated by a number of issues. First and foremost, a differential form in superspace (\textit{i.e.}, a ``superform") has significantly more components than its bosonic counterpart due to having extra legs in the fermionic directions. For example the 2-form $F$ has components
\begin{equation*}
	F = (F_{\alpha \beta}, F_{\alpha a}, F_{ab}).
\end{equation*}
The number of components is further inflated in spaces like 4D, $N = 1$ where the spinors can be chosen to be Weyl. Second, the phenomenologically interesting case of 4D, $N = 1$ has been solved in terms of unconstrained prepotential superfields and so gauge theories are often built from those instead of covariantly from the field-strengths. While not incorrect, this approach generalizes poorly to other superspaces where the prepotential solutions may be significantly more complicated or even non-existent. Additionally, an approach based on prepotentials obscures certain information that the geometry of superspace would have given you automatically.

With this in mind, we wish to show in these lectures how to take the textbook covariant construction of field-strengths---something notorious for requiring consistency checks involving highly complicated $D$-identities---and turn it into a tool that is both easy to use and generalizes effortlessly to other superspaces. The story begins in 4D, $N = 1$ with the geometric construction of the vector multiplet. This has been textbook material for quite some time and we will follow the corresponding chapter (XIII) in Wess and Bagger's (WB) well-known text \cite{Wess:1992cp} fairly closely for lecture \ref{sec:lec1}. In lecture \ref{sec:lec2} we show how drastically this approach simplifies after introducing a new ring of super-commutative variables and carefully studying the cohomology of a particular coboundary operator, following the work in \cite{Linch:2014iza, Randall:2014gza}. In lecture \ref{sec:lec3} we use this technology to easily reproduce the super-de Rham complex of $p$-form gauge fields in 4D, $N = 1$ (originally formulated by Jim Gates \cite{Gates:1980ay}) and further to trivially derive the prepotential superfields in that superspace. Lecture \ref{sec:lec4} is an application of this methodology to the construction of the abelian tensor hierarchy (ATH) in \cite{Becker:2016xgv}, while lecture \ref{sec:lec5} extends the arguments to the full non-abelian tensor hierarchy (NATH) of \cite{Becker:2016rku}. We conclude with lecture \ref{sec:lec6} wherein the Chern-Simons-type actions of the (N)ATH are derived by studying the cohomology of composite superforms.

For these lectures an elementary knowledge of superspace at the level of the coordinates, the algebra, superfields, and the super-covariant derivatives is presumed. We will also require that the reader have a somewhat firm grasp on the Weyl spinor algebra in four dimensions, at least at the level of the identities in appendix \ref{sec:ids}. For the final three lectures additional general knowledge of the non-abelian tensor hierarchy as introduced in \cite{Becker:2016rku} is assumed. It is not necessary to understand the full construction, however, and the important pieces are reviewed at the beginning of the relevant lectures. Throughout the lectures the differential geometry of superspace and all cohomological arguments are introduced slowly and developed from scratch. Each section also has an associated exercise to assist in the development of computational proficiency with the new tools. Fully worked solutions to the exercises are provided in appendix \ref{sec:sols}.

Finally, it is a great pleasure to thank William D. Linch \textsc{iii} for valuable discussions and insights on the content of these lectures and Daniel Robbins for helpful suggestions on how to improve certain explanations. Additionally, we wish to express our gratitude to Katrin and Melanie Becker along with the Mitchell Institute for their hospitality and invitation to present this work.

\vspace{10pt}
\el

\vspace*{\fill}
\newpage

\section{Gauge Theories from Superforms}
\label{sec:lec1}
\setcounter{page}{1}
\pagenumbering{arabic}

The conventional approach to building covariant abelian field-strengths is reviewed in the familiar context of 4D, $N = 1$ superspace. An explicit example is carried out for $p = 2$ with part of the calculation left as an exercise.

\subsection{Superspace Closure}

The standard flat 4D, $N = 1$ superspace\footnote{Unfortunately a full review of superspace is not possible here, although it would also be largely unnecessary. The relevant information is sketched out in the first paragraph of this section, and if any portion is unfamiliar a more detailed overview of superspace itself, along with superfields, the super-Poincar\'e algebra, and the supercovariant derivatives can be found in chapter IV of WB. Furthermore, any time we discuss 4D, $N = 1$ superspace in these notes we will be using the conventions of that text.} has four real dimensions described by commuting coordinates $x^a$ and four fermionic dimensions (corresponding to the four supercharges) described by anti-commuting coordinates $(\theta^\alpha, \bar{\theta}^\ad)$. This superspace is denoted by $\R^{4|4}$ or simply $4|4$. Functions on this space are called superfields and can famously be expanded in $\theta$ to obtain a quickly terminating series. The terms in this series then have coefficient fields that comprise a representation of the supersymmetry algebra---a supermultiplet. If the superfield is unconstrained this multiplet is highly reducible. To impose constraints covariantly on the superfields (so as to enforce irreducibility) the supercovariant derivatives $D_\alpha$ must be introduced since the standard fermionic derivatives $\partial / \partial \theta$ do not commute with the supersymmetry generators.

A reasonable question is to then ask which covariant constraints define the abelian $p$-form field-strength multiplets. The conventional answer to such a question is to formulate the notion of a differential form in superspace, then demand that the form be closed so that it must be locally exact. The superspace closure of the form will put constraints on the superfield inside the form and so the exercise then becomes one of extracting the superfield constraints from the statement of closure. Before getting too far ahead of ourselves however, let us start at the beginning and see how this works in detail.

The superspace de Rham operator is
\begin{equation}
\label{eq:super_dR}
	\d \ceq \d x^m \frac{\partial}{\partial x^m} + \d \theta^\mu \frac{\partial}{\partial \theta^\mu} + \d \bar{\theta}^\md \frac{\partial}{\partial \bar{\theta}^\md} \equiv \d z^M \partial_M
\end{equation}
for $z^M = (x^m, \theta^\mu, \bar{\theta}^\md)$. Furthermore, a superform $\omega \in \Omega^p(\R^{4|4})$ of degree $p$ is expressed in terms of the $\d z$ as
\begin{equation}
\label{eq:superform}
	\omega = \d z^{M_1} \wedge \ldots \wedge \d z^{M_p} \omega_{M_p \ldots M_1},
\end{equation}
where the $\omega_{M_p \ldots M_1}$ are the (coordinate) \textit{components} of $\omega$. These components are \textit{a priori} unconstrained superfields. The exterior derivative of $\omega$ is then
\begin{equation}
	\d \omega = \d z^{M_1} \wedge \ldots \wedge \d z^{M_p} \wedge \d z^N \partial_N \omega_{A_{M_p} \ldots A_{M_1}}.
\end{equation}
Additionally, it is clear from this formulation that $\d^2 = 0$. However, the components of $\d \omega$ are not superfields since $\partial_N$ does not commute with the supercharges. To fix this, we ``change coordinates"
\begin{equation}
	\d z^M \partial_M = (\d z^M e_M{}^A) (e_A{}^N \partial_N) \equiv e^A D_A
\end{equation}
\textit{via} the invertible matrix
\begin{equation}
	e_A{}^M = \left( \begin{array}{ccc}
		e_a{}^m & e_a{}^\mu & e_{a \md} \\
		e_\alpha{}^m & e_\alpha{}^\mu & e_{\alpha \md} \\
		e^{\ad m} & e^{\ad \mu} & e^\ad{}_\md
	\end{array} \right) = \left( \begin{array}{ccc}
 		\delta_a^m & 0 & 0 \\
 		i (\sigma^m \bar{\theta})_\alpha & \delta_\alpha^\mu & 0 \\
 		i (\theta \sigma^m \epsilon)^\ad & 0 & \delta^\ad_\md
	\end{array} \right)
\end{equation}
so that the $D_A$ become the super-covariant derivatives that commute with the supersymmetry generators $Q_\alpha$ to form the extension of the Poincar\'e algebra,
\begin{equation}
\label{eq:super_alg}
	\{D, Q\} = 0, \qquad \{D_\alpha, \bar{D}_\ad\} = T_{\alpha \ad}^a \partial_a = - 2i \partial_{\alpha \ad}.
\end{equation}
Here we have also explicitly shown the flat-superspace torsion component $T_{\alpha \ad}^a$. In this new basis the de Rham differential is now
\begin{equation}
	\d = e^A D_A
\end{equation}
and \eqref{eq:superform} becomes
\begin{equation}
	\omega = e^{A_1} \wedge \ldots \wedge e^{A_p} \omega_{A_p \ldots A_1}.
\end{equation}
The penalty for making this change becomes apparent when we try to take the the exterior derivative of $\omega$ since the basis elements are no longer killed by the differential. Instead,
\begin{equation}
	\d e^a = - 2i e^\alpha \sigma^a_{\alpha \ad} e^\ad = e^\alpha T^a_{\alpha \ad} e^\ad
\end{equation}
and so
\begin{equation}
	\d \omega = e^{A_1} \wedge \ldots e^{A_{p + 1}} (\d \omega)_{A_{p + 1} \ldots A_1}
\end{equation}
for the (frame) components
\begin{equation}
\label{eq:conv_closure}
	(\d \omega)_{A_1 \ldots A_{p + 1}} = \frac{1}{p!} D_{[A_1} \omega_{A_2 \ldots A_{p + 1}]} + \frac{1}{2! (p - 1)!} T_{[A_1 A_2|}{}^C \omega_{C |A_3 \ldots A_{p + 1}]}.
\end{equation}
Here we have used the index notation $[A_1 \ldots A_{p + 1}]$ to denote the graded index symmetrization. This means that spinor indices are symmetrized while all other index pairs are anti-symmetrized. Additionally, the only non-zero component of the torsion in flat space is the component $T_{\alpha \ad}^a$ defined in \eqref{eq:super_alg}.

In order to define an irreducible closed superform, it is the vanishing of \eqref{eq:conv_closure} that must be iteratively solved by setting to zero some of the irreducible parts in the components $\omega_{A_1 \ldots A_p}$. In practice---and we will see an explicit example in the next section---this corresponds to setting as many of the lower-dimension components to zero as possible. This allows us to solve for the lowest non-zero component in terms of a single superfield that then becomes covariantly constrained as we solve the remaining closure conditions.\footnote{The closure conditions are often also referred to as \textit{Bianchi identities}.} For concreteness, we abstain from any further discussions of the technicalities until we have an example of how the procedure works under our belts.

\subsection{Building the 4D Vector Multiplet}

Suppose we now wanted to understand the superfield structure of a supersymmetric vector multiplet in $4|4$. The field-strength $F$ of a gauge 1-form $A$ is a 2-form that we must constrain to be closed under the action of the superspace de Rham operator. The components of this form are
\begin{equation}
	F = (F_{\alpha \beta}, F_{\ad \beta}, F_{\ad \bd}, F_{\alpha a}, F_{\ad a}, F_{ab})
\end{equation}
which are graded by engineering dimension
\begin{equation}
	[F_{AB}] = (\text{\# of vector indices}) + \tfrac{1}{2}(\text{\# of spinor indices}).
\end{equation}
The closure conditions, taken directly from equation (13.20) of WB, relate these components. The dimension-$\tfrac{3}{2}$ conditions read
\begin{subequations}
\label{eq:conv_dim32_closure}
\begin{align}
	0 & = D_\gamma F_{\beta \alpha} + D_\beta F_{\alpha \gamma} + D_\alpha F_{\gamma \beta}, \\
	0 & = \bar{D}_\gd F_{\beta \alpha} + D_\beta F_{\alpha \gd} + D_\alpha F_{\gd \beta} + 2i \sigma_{\beta \gd}^a F_{a \alpha} + 2i \sigma_{\alpha \gd}^a F_{a \beta}, \\
	0 & = D_\gamma F_{\bd \ad} + \bar{D}_\bd F_{\ad \gamma} + \bar{D}_\ad F_{\gamma \bd} + 2i \sigma_{\gamma \bd}^a F_{a \ad} + 2i \sigma_{\gamma \ad}^a F_{a \bd}, \\
	0 & = \bar{D}_\gd F_{\bd \ad} + \bar{D}_\bd F_{\ad \gd} + \bar{D}_\ad F_{\gd \bd},
\end{align}
\end{subequations}
and so on for the dimension-2, dimension-$\tfrac{5}{2}$, and dimension-3 conditions.

Let us not sugarcoat things: This is a mess. There are a total of ten coupled super-differential equations that must be solved to understand the simplest gauge theory in the nicest superspace. Higher-degree gauge theories are more complicated, especially in higher dimensions or with more supersymmetry. Nonetheless, it is worthwhile to work through some of the relations to see what kind of improvements we could hope to make to the process.

First we must impose some set of initial constraints that will uniquely fix the rest of the superform structure. In WB the correct set of initial constraints is presented without explanation as
\begin{equation}
\label{eq:conv_constraints}
	F_{\alpha \beta} = F_{\ad \beta} = F_{\ad \bd} = 0.
\end{equation}
In lecture \ref{sec:lec2} we will explain why this constraint makes sense. For now we note simply that it reduces \eqref{eq:conv_dim32_closure} to
\begin{subequations}
\begin{align}
	0 & = 2i \sigma_{\beta \gd}^a F_{a \alpha} + 2i \sigma_{\alpha \gd}^a F_{a \beta}, \\
	0 & = 2i \sigma_{\gamma \bd}^a F_{a \ad} + 2i \sigma_{\gamma \ad}^a F_{a \bd}.
\end{align}
\end{subequations}
The solution to these is
\begin{equation}
\label{eq:conv_dim32_component}
	F_{\alpha a} = - i (\sigma_a)_{\alpha \ad} \bar{W}^\ad \quad \text{and} \quad F_{\ad a} = - i W^\alpha (\sigma_a)_{\alpha \ad}
\end{equation}
for the (arbitrarily normalized) spinor superfields $(W^\alpha, \bar{W}^\ad)$, related by conjugation to ensure irreducibility. Next up are the dimension-2 closure conditions. After imposing \eqref{eq:conv_constraints} they become
\begin{subequations}
\label{eq:conv_dim2_closure}
\begin{align}
	0 & = D_\beta F_{\alpha c} - D_\alpha F_{c \beta}, \\
	0 & = \bar{D}_\bd F_{\ad c} - \bar{D}_\ad F_{c \bd}, \\
\label{eq:conv_dim2_ssbv}
	0 & = \bar{D}_\bd F_{\alpha c} - D_\alpha F_{c \bd} + 2i \sigma_{\alpha \bd}^a F_{ac}.
\end{align}
\end{subequations}
Plugging in the solutions \eqref{eq:conv_dim32_component} we find---decomposing each of the identities into their irreducible parts---the top component
\begin{equation}
\label{eq:conv_dim2_component}
	F_{ab} = - \tfrac{1}{2} (\bar{D} \bar{\sigma}_{ab} \bar{W} - D \sigma_{ab} W)
\end{equation}
along with the covariant constraints
\begin{align}
\label{eq:conv_W_cons}
	\bar{D}_\ad W_\alpha = D_\alpha \bar{W}_\ad = 0 \quad \text{and} \quad D^\alpha W_\alpha - \bar{D}_\ad \bar{W}^\ad = 0.
\end{align}
It turns out that these are the only constraints on $W_\alpha$ imposed by the closure of $F$. The dimension-$\tfrac{5}{2}$ and dimension-3 closure conditions are automatically satisfied by the components we found and the constraints thereon. However, these checks are not trivial; the final condition is left as an exercise to verify.

The solution to these constraints is
\begin{equation}
	W_\alpha = - \tfrac{1}{4} \bar{D}^2 D_\alpha V
\end{equation}
for a real scalar superfield $V$. The superfield $V$ is referred to as the \textit{prepotential}. This solution is also easily verified. However, it may not be obvious how to construct this solution if it were not given to you and if you were not sufficiently experienced with the covariant derivative algebra.\footnote{A number of useful $D$-identities for $4|4$ can be found in appendix \ref{sec:ids}.} This is a legitimate issue, especially for anyone looking to study the gauge theories of a superspace where the prepotential solutions---if they exist---have not already been constructed. We will return to this point in lecture \ref{sec:lec2} and in lecture \ref{sec:lec3} show how we can trivially reconstruct the entire set of 4D, $N = 1$ prepotentials. For now, we remark that $V$ is the gauge superfield whose component fields comprise the 4D, $N = 1$ vector multiplet and $W_\alpha$ is the associated field-strength superfield invariant under the gauge transformation
\begin{equation}
	V \rightarrow V' = V + \Phi + \bar{\Phi}
\end{equation}
for any chiral field $\Phi$.\footnote{An elementary discussion of the $4|4$ vector multiplet can be found in chapter VI of WB.}

This is the traditional approach to superforms. We began by pulling a set of initial constraints \eqref{eq:conv_constraints} on the components out of hat. It was then necessary to grind through a fair bit of spinor algebra to the find the components \eqref{eq:conv_dim32_component} and \eqref{eq:conv_dim2_component} and, more importantly, the superfield constraints \eqref{eq:conv_W_cons}. The components were found in terms of a superfield field-strength $W_\alpha$ and the constraints on that superfield were solved in terms a prepotential superfield. Overall, quite a bit of work was done for a relatively simple result. Quite a bit more, perhaps, than was absolutely necessary.

With experience, the art of constructing a gauge theory from a superform is simplified slightly. The initial constraints become easier to guess and the lore is developed that the dimension-$(p + 1)$ and dimension-$(p + \tfrac{1}{2})$ closure conditions for a $p$-form never impose any new constraints on the superfield field-strength except in special cases. Unfortunately, even experienced practitioners can struggle with superform calculations in more complicated superspaces. Constraints are often difficult to isolate from terms in the closure conditions that cancel against each other or are absorbed into components. Even worse is when constraints appear at multiple dimensions in the closure conditions and new constraints must be carefully separated from relations already implied by the lower ones. In 5D, $N = 1$ superspace the closure conditions for the 3-form at dimension-3 and dimension-$\tfrac{7}{2}$ produce such a disaster of interrelated pieces that I have never been able to finish the direct check of their satisfaction.

In this lecture we have tried to orient the reader so that the ideas of the superform construction are clear while the messy calculations are merely tolerated as an unfortunate defect of carrying out the procedure in superspace. The reader is encouraged to work out as many of these calculations as they feel are necessary to frustrate themselves into longing for a better approach. In particular, the dimension-3 closure condition on $F$ is left as an exercise in checking a relation that tells you absolutely nothing new about the superform.

In the next lecture we slightly reformulate the superform construction and in doing so show how to explicitly answer the many questions that have arisen in this first lecture. In particular, we will:
\begin{enumerate}
	\item Show how to easily identify the initial superform constraints,
	\item Isolate where the superfield constraints sit inside the closure conditions,
	\item Classify which closure conditions impose new constraints, and
	\item Demonstrate how to trivially ``solve" the superfield constraints.
\end{enumerate}
The lesson to learn is that the difficulty of the conventional approach is not a deficiency of superspace. Rather, the fault is our own for failing to exploit what the geometry of superspace is trying to give us for free.

\hypertarget{ex1}{}
\subsection{Exercise \#1}

Verify that the dimension-3 closure condition
\begin{equation}
	\partial_{[a} F_{bc]} = 0
\end{equation}
follows from the components \eqref{eq:conv_dim32_component} and \eqref{eq:conv_dim2_component} and the constraints \eqref{eq:conv_W_cons}, or, more easily, from the closure conditions that imply them. You should assume that you have already proven the automatic satisfaction of dimension-$\tfrac{5}{2}$ conditions from these same components and constraints. Appendix \ref{sec:ids} will prove to be a useful reference for this and the remaining exercises. (\hyperlink{sol1}{\textbf{Solution \#1}})

\vspace{10pt}
\el
\clearpage

\section{Closure from Lie Algebra Cocycles}
\label{sec:lec2}

A new approach to constructing covariant field-strengths is presented based the nilpotency of the super-de Rham operator. The geometric closure conditions of lecture \ref{sec:lec1} are solved after calculating the appropriate Lie algebra cocycles for a coboundary operator constructed from the flat-superspace torsion. The derivation of these cocycles for $4|4$ is left as an exercise.

\subsection{Consequences of Nilpotency}

Let us begin by proposing a modest notational change to the formalism of lecture \ref{sec:lec1}. The graded anti-symmetry in the indices of the components $\omega_{A_1 \ldots A_p}$ led to closure conditions like \eqref{eq:conv_dim32_closure} with multiple terms of the same type. One can imagine this quickly becoming unwieldy for higher-degree forms. At this point we must note that in this lecture we will be working in an arbitrary superspace $m|n$ with abstract (pseudo)real spinors $\lambda^\alpha$. This notation can be made precise upon specializing to a particular superspace and over the course of this and the following lecture everything will be shown explicitly for $4|4$.

For computational clarity, we now introduce the constants $j^A = (\psi^a, s^\alpha)$ with the commutation relations
\begin{equation}
	[s^\alpha, j^A] = 0 \quad \text{and} \quad \{\psi^a, \psi^b\} = 0.
\end{equation}
We will also make heavy use of the geometric notation $j^A X_A \eqc X_j$. If we contract $j^{A_1} \ldots j^{A_p}$ onto \eqref{eq:conv_closure} we obtain the simplified closure condition
\begin{align}
\label{eq:abs_closure}
	0 & = s D_s \omega_\janksubp + (-1)^s (p + 1 - s) \partial_\psi \omega_\janksubp \notag\\
		& ~~ - \tfrac{1}{2} (-1)^s s (s - 1) T_{ss}^a \omega_\janksub{a},
\end{align}
where $s$ is the number of $s^\alpha$ contractions on each term in the formula, $p$ is the degree of $\omega$, and $T_{\alpha \beta}^a$ is the flat-superspace torsion. The coefficient on the first term is due to the $s$ different ways that one of the contracted $s^\alpha$s can end up on the $D$ instead of the form component. The coefficient of the second term comes from the $(p + 1 - s)$ ways that a $\psi^a$ can end up on the $\partial$ and a potential sign from having to rearrange indices using $\omega_{\ldots a \alpha \ldots} = - \omega_{\ldots \alpha a ...}$. The coefficient on the torsion term involves the same reasoning about the sign and takes into account the $s (s - 1)$ ways that two $s^\alpha$s get contracted onto the torsion.

In $4|4$ our spinor $\lambda^\alpha$ splits into $\lambda^\alpha \rightarrow \lambda^\alpha \oplus \bar{\lambda}^\ad$ to accommodate both the fundamental and anti-fundamental representations of $SL(2, \mathbf{C})$. The specific form of \eqref{eq:abs_closure} in $4|4$ then expands to
\begin{align}
\label{eq:44_closure}
	0 & = s D_s \omega_\janksubfp + \sb \bar{D}_\sb \omega_\janksubfp \notag\\
			& \quad + (-1)^{s + \sb} (p + 1 - s - \sb) \partial_\psi \omega_\janksubfp \notag\\
			& \quad + 2i (-1)^{s + \sb} s \sb \omega_\janksubfs.
\end{align}
Here $s$ (or $\sb$) is the number of $s^\alpha$ (or $\sb^\ad$) contractions, $p$ is again the degree of $\omega$, and we have also introduced the vector
\begin{equation}
	\sigma^a(s, \sb) \ceq s^\alpha \sigma^a_{\alpha \ad} \sb^\ad
\end{equation}
so that
\begin{equation}
	T_{s \sb}^a = - 2i \sigma^a(s, \sb).
\end{equation}
This we contracted on the form component as $T_{s \sb}^a \omega_{\ldots a \ldots} = - 2i \omega_{\ldots \sigma(s, \sb) \ldots}$. It is important to point out here that the constant spinors $(s^\alpha, \sb^\ad)$ contract onto equations as they are written, without any index raising or lowering. Additionally, it is not difficult to check that \eqref{eq:44_closure} reproduces the closure conditions we solved in lecture \ref{sec:lec1} if the $j^A$ are contracted onto the latter.

Returning to $m|n$, we define the torsion by
\begin{equation}
	T_{ss}^a = - 2i \gamma^a(s, s).
\end{equation}
It is easy to verify that twice iterating \eqref{eq:abs_closure} vanishes identically; this is the statement that the super-de Rham operator $\d$ is nilpotent of order two. Indeed, since $\partial_\psi \partial_\psi = \gamma_\psi(s, s) \gamma_\psi(s, s) = 0$,
\begin{align}
	0 & \equiv s (s - 1) D_s^2 \omega_\janksubp - (-1)^s s (p + 1 - s) D_s \partial_\psi \omega_\janksubp \notag\\
		& \quad - i (-1)^s s (s - 1)(s - 2) D_s \omega_\janksubg + (-1)^s s (p + 1 - s) \partial_\psi D_s \omega_\janksubp \notag\\
		& \quad + i s (s - 1)(p + 1 - s) \partial_\psi \omega_\janksubg + i (-1)^s s (s - 1)(s - 2) D_s \omega_\janksubg \notag\\
		& \quad + i s (s - 1) \partial_{\gamma(s, s)} \omega_\janksubp - i s (s - 1)(p + 1 - s) \partial_\psi \omega_\janksubg
\end{align}
due to the super-Poincar\'e commutation relations\footnote{This suggests that the story changes in curved superspaces. It turns out there that $\d^2 \equiv 0$ is what \textit{defines} a consistent conformal supergravity. Once the supergravity construction is complete, the cohomology arguments of this lecture can be applied with surprisingly little modification.}
\begin{equation}
	D_s^2 + i \partial_{\gamma(s, s)} = 0 \quad \text{and} \quad [D_s, \partial_\psi] = 0.
\end{equation}
Furthermore, notice that in \eqref{eq:abs_closure} the component of $\omega$ in the first term has engineering dimension $p - \tfrac{1}{2} s + \tfrac{1}{2}$, the component in the second has $p - \tfrac{1}{2} s$, and the component in the third has the highest at $p - \tfrac{1}{2} s + 1$.

To reiterate, we know from our setup of the closure conditions that
\begin{align}
\label{eq:abs_d_closure}
	0 & \equiv s D_s (\d \omega)_\janksubp + (-1)^s (p + 2 - s) \partial_\psi (\d \omega)_\janksubp \notag\\
		& ~~ + i (-1)^s s (s - 1) (\d \omega)_\janksubg.
\end{align}
But how does this help us solve $(\d \omega)_\janksubp = 0$? Suppose we begin by
solving the dimension-$\tfrac{1}{2}(p + 1)$ closure condition
\begin{equation}
	(\d \omega)_{s \ldots s} = 0.
\end{equation}
In $4|4$ with $p = 2$ this corresponded to solving for $F_{\alpha a}$ and $F_{\ad a}$ in terms of a spinor superfield. This is always the easiest condition to solve and is often completely trivial. By an inductive argument, assume we have also solved the closure conditions up through dimension-$\ell$ for $\ell < p + 1$. From \eqref{eq:abs_d_closure} we then know that the dimension-$(\ell + \tfrac{1}{2})$ closure condition is constrained to satisfy
\begin{equation}
\label{eq:abs_kerdel}
	(\d \omega)_\janksubg \equiv 0.
\end{equation}
This is the most important equation in the lecture series. What we are being told here is that any \textit{new} information to be obtained from the dimension-$(\ell + \tfrac{1}{2})$ closure condition must sit in the kernel of the contraction operator
\begin{equation}
	\delta \ceq \iota_{\gamma(s, s)},
\end{equation}
which acts as $\delta \omega_\janksubp \ceq \omega_\janksubg$; the rest must drop out according to $\d^2 = 0$. The most obvious part of $\ker \delta$ is anything in $\im \delta$ since $\delta^2 = 0$. From the closure condition \eqref{eq:abs_closure} we see that these pieces are the components of the form. This means that the rest, $\ker \delta / \im \delta$, holds the constraints. The parts of the dimension-$(\ell + \tfrac{1}{2})$ condition that are not annihilated by $\delta$ must be set to zero by the solution to the lower-dimension closure conditions.

This is starting to sound promising. By using the nilpotency of the super-de Rham differential we can identify which irreducible parts of the closure condition form the constraints, which form the components, and which tell us nothing. If we could classify the elements of $\ker \delta / \im \delta$ then we might further be able to see when an entire closure condition is implied by constraints and components already derived.

\subsection{Cocycles, Coboundaries, and Principality}

The ideas of the previous section are more naturally described in the language of algebraic cohomology. In fact, the insight here is that the \textit{geometric} analysis required for a closed superform can be entirely reduced to a far simpler \textit{algebraic} cohomology problem. In this setting, the operator $\delta$ is the BRST \textit{coboundary operator}\footnote{The full BRST operator also includes the (super-)translation generators but these are precisely the parts we do not need for our analysis.} on a cochain complex graded by total ghost number for the ghosts $j^A$. The ghosts are closed ($\d j^A = 0$; torsion-free) as required by their being constant. The components of a closed superform are the \textit{coboundaries} (elements of $\im \delta$) while the constraints on the form's superfield are elements of the cohomology group $H_\delta \equiv \ker \delta / \im \delta$. Elements of $\ker \delta$ are known as the \textit{cocycles}. This terminology can help orient us mathematically (and perhaps confuse us physically; nothing is being gauge-fixed here) but the full technical machinery of cohomological algebra is not necessary.\footnote{The cohomology of the super-Poincar\'e algebra has also been studied by mathematicians, most relevantly to our purposes in \cite{Brandt:2009xv, Movshev:2010mf}. However the comprehensive cohomological calculations in those references are overkill for us; we are after something much simpler. Furthermore, the use of language concerning ``BRST operators" and ``ghosts" in those references---although we have borrowed such language here---can be slightly misleading since physically this has nothing to do with the usual BRST prescription.}

What we wish to know are the elements of $H_\delta$. For the most part the non-trivial cocycles will suffice, up to a caveat discussed in solution \hyperlink{sol2}{2}. First,
\begin{equation}
\label{eq:principal_cocycle}
	\gamma_\psi(s, \xi) \in H_\delta
\end{equation}
for an arbitrary spinor $\xi$. This element is famously responsible for the existence of supersymmetric Yang-Mills theories in $m = 3, 4, 6, 10$. In fact, this element---the \textit{principal cocycle}---is so important that we will define those superspaces ($3|2$, $4|4$, $6|8$, and $10|16$) in which a spinor representation exists so that \eqref{eq:principal_cocycle} holds as \textit{principal superspaces}. The remaining cocycles in a principal superspace can then be found by taking $\xi$ to be an arbitrary product of anti-symmetrized $\gamma$-matrices and the spinor ghosts. Carrying out this classification for $4|4$ is left as an exercise. In $6|8$ (6D, $N = (1, 0)$), where the spinors are eight-component pseudoreal objects $\lambda^{\alpha i}$, the cocycles are
\begin{equation}
	H_\delta^{6|8} = \{1, \gamma_\psi(s, \xi), \gamma_{\psi \psi \psi}(s, s)\}
\end{equation}
where $1$ denotes the scalar cocycle.

Fortunately there are a very small number of independent non-trivial cocycles in a given superspace. Principal superspaces have at most two and all other superspaces have one.\footnote{The principal cocycle \eqref{eq:principal_cocycle} fails to be a cocycle under dimensional reduction from $m|n$ to its embedded subspace $(m - 1)|n$; an example of this can be seen in appendix A of \cite{Linch:2014iza}.} From this classification it is trivial to prove the following lemma, originally given in \cite{Randall:2014gza}:
\vspace{4pt}

\hypertarget{lemma1}{}
\textbf{Lemma 1}: \textit{In a principal superspace, the set of constraints imposed on $p$-forms by the top two closure conditions is trivial for $p > 1$. For $p = 1$ the constraint $C_\alpha = 0$ sits in the closure condition as $(\d A)_{s \psi} = \gamma_\psi(s, C)$. In a non-principal superspace, the set of constraints imposed on $p$-forms by the top two closure conditions is trivial for all $p$.}
\vspace{4pt}

\textbf{Proof}: In a principal superspace the principal cocycle requires only a single $s$-contraction and so it can support a constraint term in the dimension-$\tfrac{3}{2}$ closure condition of a 1-form. For higher-degree forms the $s = 1$ closure condition will have extra $\psi$-contractions for $\delta$ to attach itself to instead of the cocycle. In a non-principal superspace the non-trivial cocycle requires two $s$-contractions since it reduces from the other cocycle in a principal superspace with the same number of supercharges. Therefore, an $s = 1$ closure condition in a non-principal superspace can never support a constraint term for any degree superform. \leafNE
\vspace{4pt}

We have now completely trivialized the process of finding the constraints on a closed superform. It may not yet seem clear how to use this in practice, but in lecture \ref{sec:lec3} we will apply it to every $4|4$ form we can get our hands on.

Thus far we have said precious little about the components of the form other than that they comprise the $\im \delta$ portion of $\d \omega$. In lecture \ref{sec:lec1} we made note of the mysterious way in which the initial constraints \eqref{eq:conv_constraints} were chosen. To shed some light on this procedure we will begin by looking at the super-de Rham complex $\Omega^\bullet(m|n)$ as a whole. Furthermore, let us suppose that $m|n$ is principal, as $4|4$ is. In this complex we have closed $p$-form field-strengths for $p = 1, 2, \ldots, m$. From lemma \hyperlink{lemma1}{1} we know that a closed 1-form $A$ will have a constraint on the superfield inside $A$ that has a single free spinor index. What would be the easiest way to build a closed 2-form $F$ with this information?

The answer is to simply obstruct the closure of $A$ as $F = \d A$. This guarantees that a consistent set of constraints on $F$ exists so that $\d F = 0$. In components, this says that
\begin{subequations}
\begin{align}
	(\d A)_{ss} & = F_{ss} = 0, \\
	(\d A)_{s \psi} & = \gamma_\psi(s, C) = F_{s \psi}, \\
	(\d A)_{\psi \psi} & = F_{\psi \psi}.
\end{align}
\end{subequations}
Because the first superfield constraint on $A$ does not appear until we reach the dimension-$\tfrac{3}{2}$ closure condition, \textit{there is nothing for the dimension-1 components of $F$ to obstruct.} So we let them vanish and then further pick the lowest non-vanishing component of the superform to be proportional to a cocycle! Then $F_{\psi \psi} \neq 0$ as well because $(\d A)_{\psi \psi}$ vanishes only when $C_\alpha = 0$. So not only do the cocycles define the constraints but they also show us where the lowest non-vanishing component of each superform is, provided that the absolute lowest component $\omega_{s \ldots s}$ of the form vanishes. The 1-forms will have a slightly different treatment because of this caveat, as will certain forms in non-principal superspaces. However, these cases are always simple to deal with since they only appear at low degrees in the super-de Rham complex.

We have gone quite far in this lecture in a fairly abstract setting. The benefit of such abstractness, however, came in our ability to draw very general conclusions without specializing to a particular superspace. To summarize the main points before jumping into $4|4$, we conclude by addressing the questions posed at the end of lecture \ref{sec:lec1}. That is, how do we:
\begin{itemize}
	\item[1.] \textit{Show how to easily identify the initial superform constraints?}
\end{itemize}
If we want to demand the closure of a $p$-form we can simply set as many of the lower components to zero until we reach one that can be written as proportional to a cocycle. In the few low-degree cases where this does not work, the superfield structure must be figured out manually (in the case of $p = 1$) or by looking directly at the index structure of the constraints on a closed $(p - 1)$-form (as in the case of $p = 2$ for $5|8$).
\begin{itemize}
	\item[2.] \textit{Isolate where the superfield constraints sit inside the closure conditions?}
\end{itemize}
To do this we simply read off which irreducible parts are proportional to the relevant cocycle. We will get a lot of practice doing this in lecture \ref{sec:lec3}.
\begin{itemize}
	\item[3.] \textit{Classify which closure conditions impose new constraints?}
\end{itemize}
This was the content of lemma \hyperlink{lemma1}{1}, where we proved that the $s = 0, 1$ closure conditions are (almost) always automatically satisfied and explained that the sole exceptions occur for $p = 1$ in principal superspaces.
\begin{itemize}
	\item[4.] \textit{Demonstrate how to trivially solve the superfield constraints?}
\end{itemize}
In the interest of keeping this lecture from growing too long, we will continue to postpone the answer to this question until the second half of lecture \ref{sec:lec3} where it will be addressed comprehensively with our full arsenal of examples in $4|4$.

\hypertarget{ex2}{}
\subsection{Exercise \#2}

Compute the cocycles (more precisely, the elements of $H_\delta$) for the 4D, $N = 1$ super-Poincar\'e algebra. Use the fact that $4|4$ is a principal superspace and choose $\xi$ appropriately. (\hyperlink{sol2}{\textbf{Solution \#2}})

\vspace{10pt}
\el
\clearpage

\section{4D, $N = 1$ Super-de Rham Complex}
\label{sec:lec3}

The super-de Rham complex in $4|4$ is derived from scratch and prepotential solutions constructed from the nilpotency of the super-de Rham differential. The investigation of an additional, little-studied superform is left as an exercise to practice with the efficiency of the new approach.

\subsection{Superfield Constraints}

In this lecture we will put to use the abstract formalism of lecture \ref{sec:lec2} for the purpose of constructing the super-de Rham complex of $p$-form field-strengths in $4|4$. The key equations necessary for us to do so are the closure condition \eqref{eq:44_closure} and the cocycles \eqref{eq:44_cocycles} worked out in exercise \hyperlink{ex2}{2}.

We begin, as is often done, at the beginning. Consider a 1-form $A \in \Omega^1(4|4)$ that we will require to be closed. The components of $A$ are $A = (A_s, A_\sb, A_\psi)$ that are \textit{a priori} unrelated. The dimension-1 closure conditions are
\begin{subequations}
\begin{align}
\label{eq:dA_ss}
	0 & = (\d A)_{ss} = 2 D_s A_s, \\
\label{eq:dA_ssb}
	0 & = (\d A)_{s \sb} = D_s A_\sb + \bar{D}_\sb A_s + 2i A_{\sigma(s, \sb)}, \\
\label{eq:dA_sbsb}
	0 & = (\d A)_{\sb \sb} = 2 \bar{D}_\sb A_\sb.
\end{align}
\end{subequations}
The first of these is satisfied when $A_s = D_s U$ and the third when $A_\sb = \bar{D}_\sb U'$ for scalar superfields $U$ and $U'$. This leaves an ambiguity (more precisely, a gauge freedom) in how we solve the rest of the constraints. The simplest solution, $U' = U$ with $A_A = D_A U$ is pure gauge (unconstrained). To have $A$ be a field-strength we must settle for the next simplest solution: $U' = - U$. Plugging this into the second constraint \eqref{eq:dA_ssb} yields
\begin{equation}
	0 = [\bar{D}_\sb, D_s] U + 2i A_{\sigma(s, \sb)} ~~\Rightarrow~~ A_\psi = - \tfrac{i}{4} \bar{\sigma}_\psi^{\alpha \ad} [\bar{D}_\ad, D_\alpha] U.
\end{equation}
The dimension-$\tfrac{3}{2}$ closure conditions are
\begin{subequations}
\begin{align}
\label{eq:dA_sv}
	0 & = (\d A)_{s \psi} = D_s A_\psi - \partial_\psi A_s, \\
\label{eq:dA_sbv}
	0 & = (\d A)_{\sb \psi} = \bar{D}_\sb A_\psi - \partial_\psi A_\sb, 
\end{align}
\end{subequations}
These are conjugate to one another so we will focus exclusively on \eqref{eq:dA_sbv} for simplicity. From the story in lecture \ref{sec:lec2} we know that the constraint sits behind the $\sigma_\psi(\cdot, \sb)$ cocycle and that any remaining terms must cancel against each other. Then the $\partial_\psi$ term will obviously contribute nothing to the constraint. We can ``pull out" the primary cocycle component of the first term as
\begin{equation}
	\bar{D}_\sb A_\psi = - \tfrac{i}{4} \bar{D}_\sb \sigma_\psi^{\alpha \ad} [\bar{D}_\ad, D_\alpha] U = - \tfrac{i}{4} \sigma_\psi(\bar{D}^2 D U, \sb) + \ldots,
\end{equation}
where $\ldots$ denotes the term due to the anti-commutator that will cancel the $\partial_\psi$ term we neglected in \eqref{eq:dA_sbv}. It is easy to check how that happens in this case, but for higher forms it becomes incredibly useful to not have to worry about how remaining pieces in the closure conditions vanish. In any case, the object sitting behind the cocycle is the constraint,
\begin{equation}
\label{eq:44_u_cons}
	\bar{D}^2 D_\alpha U = 0.
\end{equation}
By lemma \hyperlink{lemma1}{1} the dimension-2 closure conditions tell us nothing new about $U$ and so we are already done. A 1-form field-strength in $4|4$ consists of a real scalar superfield $U$ subject to the covariant constraint \eqref{eq:44_u_cons}.

To (re)construct the closed 2-form $F$ we now obstruct the closure of $A$ as $F = \d A$. As we saw at the end of lecture \ref{sec:lec2} this construction tells us that $F_{ss} = F_{s \sb} = F_{\sb \sb} = 0$ and
\begin{equation}
	F_{\sb \psi} = - i \sigma_\psi(W, \sb).
\end{equation}
for a spinor superfield $W_\alpha$. It is important that we not take the obstruction relationship $F = \d A$ too seriously. If $W_\alpha$ was literally proportional to $\bar{D}^2 D_\alpha U$ then the closure conditions on $F$ would all be trivially satisfied by $\d^2 = 0$. Instead we simply use the obstruction as a starting point, allowing it to tell us which components of $F$ should be zero. However, we will return to the idea of automatically solved closure conditions in the second half of the lecture.

The dimension-2 closure conditions on $F$ are
\begin{subequations}
\begin{align}
	0 & = (\d F)_{ss \psi} = 2 D_s F_{s \psi}, \\
	0 & = (\d F)_{s \sb \psi} = D_s F_{\sb \psi} + \bar{D}_\sb F_{s \psi} + 2i F_{\sigma(s, \sb) \psi}, \\
	0 & = (\d F)_{\sb \sb \psi} = 2 \bar{D}_\sb F_{\sb \psi}.
\end{align}
\end{subequations}
The first and third of these are trivially satisfied by demanding that $W_\alpha$ be chiral. In the second we need to pull out the cocycle $\sigma_\psi(s, \sb)$ and will not care at all about the form of $F_{\psi \psi}$. Doing so yields
\begin{align}
	0 & = - i D_s \sigma_\psi(W, \sb) - i \bar{D}_\sb \sigma_\psi(s, \bar{W}) + 2i F_{\sigma(s, \sb) \psi} \notag\\
		& = \tfrac{i}{2} \sigma_\psi(s, \sb) (D^\alpha W_\alpha - \bar{D}_\ad \bar{W}^\ad).
\end{align}
Try carrying out this calculation for yourself. In the first term, pulling the $s$ off the $D$ and contracting it on the Pauli matrix requires anti-symmetrizing the spinor indices on $D$ and $W$. The symmetric part is then discarded (and will end up canceling the $F_{\sigma(s, \sb) \psi}$ term). It is important to not care about the parts that do not sit behind cocycles; one example why is seen in the exercise for this lecture where we run into the issue of multiple constraints at different engineering dimensions.

Thus, we have the 2-form field-strength superfield $W_\alpha$ which is constrained to be chiral and to satisfy
\begin{equation}
\label{eq:44_w_cons}
	D^\alpha W_\alpha = \bar{D}_\ad \bar{W}^\ad.
\end{equation}
Continuing up the complex, let us obstruct the closure of $F$ as $H = \d F$. There is an ambiguity here that is, as far as we are aware, entirely glossed over in the literature on $\Omega^\bullet(4|4)$. Which constraint do we obstruct, chirality or \eqref{eq:44_w_cons}? It turns out that obstructing \eqref{eq:44_w_cons} will give the tensor multiplet, but we leave the analysis of the other obstruction to be completed as an exercise. Obstruction of \eqref{eq:44_w_cons} generates the 3-form $H$ with
\begin{equation}
	H_{sss} = H_{ss \sb} = H_{s \sb \sb} = H_{\sb \sb \sb} = H_{ss \psi} = H_{\sb \sb \psi} = 0
\end{equation}
and
\begin{equation}
\label{eq:dR_H_ssbv_comp}
	H_{s \sb \psi} = - \tfrac{1}{2} \sigma_\psi(s, \sb) H
\end{equation}
for a real scalar superfield $H$.\footnote{The normalization of $H$ is such that when used in other contexts it will reproduce certain results exactly. The same will be true for the 4-form.} At this point in the de Rham complex the primary cocycle is largely irrelevant for the determination of constraints. The dimension-$\tfrac{5}{2}$ closure condition, for instance, ends up doing nothing other than defining the next components, $H_{s \psi \psi}$ and $H_{\sb \psi \psi}$. This may be confusing since
\begin{equation}
	0 = (\d H)_{s \sb \sb \psi} = 2 \bar{D}_\sb H_{s \sb \psi} - 4i H_{\sb \sigma(s, \sb) \psi}
\end{equation}
appears as though it could support a constraint term. However, as noted in solution \hyperlink{sol2}{2}, there is no element in cohomology that can be pulled out of this condition given \eqref{eq:coh_to_im_el} and the structure of \eqref{eq:dR_H_ssbv_comp}. Instead, we find
\begin{equation}
	H_{\sb \psi \psi} = - \tfrac{i}{2} \bar{\sigma}_{\psi \psi}(\sb, \bar{D}) H.
\end{equation}
The dimension-3 closure condition yields the first non-trivial cohomology as
\begin{equation}
	0 = (\d H)_{\sb \sb \psi \psi} = 2 \bar{D}_\sb H_{\sb \psi \psi} = - \tfrac{i}{2} \bar{\sigma}_{\psi \psi}(\sb, \sb) \bar{D}^2 H,
\end{equation}
which states that $H$ is \textit{linear}; that is, $H \in \ker \bar{D}^2$. From this the remaining higher-dimension closure conditions are identically satisfied.

Finally, we turn to the 4-form which is unsurprisingly generated as $G = \d H$. Such an obstruction gives rise to a complex scalar superfield $G$ and the lowest non-vanishing component
\begin{equation}
	G_{\sb \sb \psi \psi} = - 2i \bar{\sigma}_{\psi \psi}(\sb, \sb) G.
\end{equation}
Similarly to the 2-form, this superform is constrained by chirality in the dimension-$\tfrac{7}{2}$ closure condition although this is the sole constraint at that level. Before being walled off by lemma \hyperlink{lemma1}{1} we may still hope to find an additional constraint in the dimension-4 closure conditions. Alas, there we run into something familiar from the 3-form analysis; the only possibility for a constraint at that level would require that $G_{\psi \psi \psi \psi} \in \ker \delta$ and again no such cocycle exists.

All together, the complex $\Omega^\bullet(4|4)$ is built from the field-strength superfields
\begin{equation}
	(U, W_\alpha, H, G) = (\text{real + \eqref{eq:44_u_cons}, chiral + \eqref{eq:44_w_cons}, real + linear, chiral}).
\end{equation}
Let us now solve these constraints in terms of the prepotential superfields.

\subsection{Prepotential Solutions}

The complex $\Omega^\bullet(4|4)$ was originally presented in \cite{Gates:1980ay}. There it was noted how similar the constraint solutions are to the gauge variations of the prepotentials. Though not realized at the time, this is simply a manifestation of $\d^2 = 0$. Furthermore, the prepotential solutions themselves can also be derived from this nilpotency; if you are sufficiently familiar with $4|4$ you may have guessed how this works, especially since we already alluded to it.

Consider the closed 4-form $\d G = 0$. In terms of superfields, this says that
\begin{equation}
	\bar{D}_\ad G = 0.
\end{equation}
The easiest solution to $\d G = 0$ is $G = \d X$. At the level of superfields,
\begin{equation}
	G = \bar{D}^2 X
\end{equation}
for an arbitrary superfield $X$. This is the prepotential for $G$. We have switched notation from $H$ to $X$ to remind ourselves that $X$ is neither linear nor real. And that's it---all we have to do to find a prepotential solution is plug in the constraint we obstructed. Then $\d^2 = 0$ guarantees that we have a solution. Continuing downward we have the linearity of $H$; this is solved by
\begin{equation}
	H = D^\alpha \Sigma_\alpha - \bar{D}_\ad \bar{\Sigma}^\ad
\end{equation}
where $\Sigma$ is a chiral (since that part of $\d F = 0$ was not obstructed in this case) prepotential. The scalar constraint on $W$ is solved by
\begin{equation}
	W_\alpha = - \tfrac{1}{4} \bar{D}^2 D_\alpha V
\end{equation}
for a real scalar prepotential $V$. The normalizations of the prepotentials are irrelevant and merely conventional. Finally, we end up at the constraint \eqref{eq:44_u_cons}. Since the closure of the 1-form did not come from any kind of obstruction, this condition must be solved by hand. We emphasize that this is the only solution that does not come for free.\footnote{You will see in exercise \hyperlink{ex4}{4} that the conclusion is even nicer. For more complicated complexes, like those of the (N)ATH, the prepotential solutions are now \textit{entirely} trivialized after finding this bottom solution in the standard complex.} The solution is easily checked to be
\begin{equation}
	U = \tfrac{1}{2} (\Phi + \bar{\Phi})
\end{equation}
for a chiral prepotential $\Phi$.

As a technical aside, we should note that solving the obstructed constraints is not exactly equivalent to constructing a set of prepotentials. A prepotential for a given field-strength is a completely unconstrained superfield in terms of which the field-strength can be expressed so as to solve all constraints on the field-strength. In $4|4$ we can be slightly imprecise and talk about chiral or real prepotentials simply because those constraints are so easily solved in terms of true prepotentials. For example, by writing
\begin{equation}
	H = D^\alpha \Sigma_\alpha - \bar{D}_\ad \bar{\Sigma}^\ad = D^\alpha \bar{D}^2 \sigma_\alpha - \bar{D}_\ad D^2 \bar{\sigma}^\ad
\end{equation}
we see that the true prepotential for $H$ is a complex spinor $\sigma$. But this is overly messy for simply solving the linearity condition, so $\Sigma$ is termed the prepotential for $H$ with the implicit understanding that the chirality condition on $\Sigma$ can itself be easily solved. In higher-dimensional superspaces, this feature disappears. In $6|8$, for example, the 2-form field-strength also has two constraints, one of which is a scalar constraint.\footnote{For additional details on the super-de Rham complex in $6|8$, see \cite{Arias:2014ona}.} This is obstructed to get the 3-form, and so the ``prepotential" superfield for the 3-form is easily written down. However, this superfield is also subject to an additional condition whose solution is not found as effortlessly as in the case of chirality. In fact, a fully geometric understanding of the gauge prepotentials in eight-supercharge superspaces  ends up requiring the use of harmonic superspace \cite{Galperin:2001uw}. The situation is worse with more supercharges as off-shell sets of prepotentials do not even universally exist. So when we claim to be solving the covariant constraints, it is important to keep in mind that we are only solving what $\d^2 = 0$ allows us to solve. The parts of a complex that involve only a single obstructed constraint are completely solved from nilpotency, while forms with multiple constraints can end up derailing our attempts to construct prepotential solutions.

Returning to considerations of $4|4$, the gauge variations come out in an equally simple way. Starting again at the top of the complex, the exact form $G = \d X$ is invariant under $\delta X = \d \lambda$. In terms of superfields, this is
\begin{equation}
	\delta X = D^\alpha \lambda_\alpha - \bar{D}_\ad \bar{\lambda}^\ad
\end{equation}
for a chiral spinor $\lambda$, due to the constraint structure of a closed 2-form. Working our way down we find further that
\begin{align}
	\delta \Sigma_\alpha & = - \tfrac{i}{4} \bar{D}^2 D_\alpha L, \\
	\delta V & = - \tfrac{1}{2} (\Lambda + \bar{\Lambda}), \\
	\delta \Phi & = 0,
\end{align}
for $L$ real and $\Lambda$ chiral. Again, the normalizations of these fields are irrelevant.

Thus ends our time in $4|4$. From nothing more than $\d^2 = 0$ (and one small calculation for $U$) we have reconstructed the entire set of $4|4$ prepotentials and their gauge transformations. The complex of field-strengths came out almost as simply, following from $\d^2 = 0$, the classification of cocycles, and some spinor algebra. In the following lectures we will begin to look at how the ideas and methods of these first lectures can be used to simplify the structure of the non-abelian tensor hierarchy in \cite{Becker:2016rku}. In doing so we leave behind the niceties of $4|4$ and enter territory in which brute-force methods produce calculations more daunting than those encountered in lecture \ref{sec:lec1}. However, by letting the nilpotency of various ``extended" differentials guide us we will find that the superfield constraints, prepotential solutions, and gauge transformations can be derived with no more difficulty than in $4|4$.

\hypertarget{ex3}{}
\subsection{Exercise \#3}

Instead of choosing $W_\alpha$ to be chiral and obstructing \eqref{eq:44_w_cons} with the scalar $H$, obstruct the chirality condition with a vector superfield $H_a$ sitting inside a superform $H'$. Write down the lowest component of $H'$ and derive the dimension-3 constraint on $H_a$ using the fact that
\begin{equation}
\label{eq:hprime_svv}
	H'_{s \psi \psi} = - \tfrac{i}{8} \sigma_\psi(s, \bar{D}) H_\psi,
\end{equation}
provided the coefficient of the lowest component is unity. Be careful to not start calculating too much; the only thing you are after is the term sitting behind the relevant cocycle. Does this give rise to a different 4-form when obstructed or does the complex ``re-join" at $G$? (\hyperlink{sol3}{\textbf{Solution \#3}})

\textbf{Note}: The multiplet described by $H_a$ consists of gauge parameters for the conformal graviton \cite{Linch:2014iza}. This kind of thing is avoidable in 4D with a chiral $W_\alpha$ but becomes a generic part of any super-de Rham complex for $D > 4$. 

\vspace{10pt}
\el
\clearpage

\section{Extending the Complex to the ATH}
\label{sec:lec4}

The super-de Rham complex of lecture \ref{sec:lec3} is extended by an internal derivative $\partial$ that is motivated from the compactification of a higher-dimensional theory with gauge fields. Constraints are derived from the closure of the extended differential and are shown to be equivalent to the consistency conditions of the abelian tensor hierarchy. Computations involving the prepotentials, including their derivation, are left as an exercise.

\subsection{Superfield Constraints}

In this lecture we will show how the superfield constraint equations of the abelian tensor hierarchy in \cite{Becker:2016xgv} arise from superform closure under a differential $Q$. This differential consists of the usual de Rham operator plus an additional component $\partial$ that can be thought of as a derivative on an internal space $M$. Here we will take $M$ to be seven-dimensional, although it could (with minimal modifications to what follows), in principle, have some other dimension. The complex $\Omega^\bullet(\mathbf{R}^{4|4} \times M)$ then has elements graded by bi-degree $(p, q)$. From a four-dimensional point of view, we will consider ${\bm \omega} \in \Omega^4(\mathbf{R}^{4|4} \times M)$. In terms of spacetime degree $p$ the possible choices of bi-degree for ${\bm \omega}$ are
\begin{equation}
	{\bm \omega} = \left\{ \begin{array}{c}
		E, ~\text{with}\, \bdeg(E) = (0, 4), \\
		A, ~\text{with}\, \bdeg(A) = (1, 3), \\
		F, ~\text{with}\, \bdeg(F) = (2, 2), \\
		H, ~\text{with}\, \bdeg(H) = (3, 1), \\
		G, ~\text{with}\, \bdeg(G) = (4, 0).
	\end{array} \right.
\end{equation}
Denote any of these by $\omega_p$ with bi-degree $(p, 4 - p)$. The differential $Q$ acts as
\begin{equation}
	Q \omega_p \ceq \d \omega_p + (-1)^{p + 1} \partial \omega_{p + 1}.
\end{equation}
Since $\bdeg(\d \omega_p) = (p + 1, 4 - p)$ and $\bdeg(\partial \omega_p) = (p, 4 - p + 1)$ we have
\begin{equation}
	\bdeg(Q \omega_p) = (p + 1, 4 - p).
\end{equation}
Action by $Q$ increases total degree by one. Furthermore, since closure of ${\bm \omega}$ does not depend on the internal degree $q$ in any meaningful way we will still be able to use iterative obstructions to generate the complex as we did in $4|4$. This will of course be important for building prepotential solutions to whatever constraints we find. Finally, the nilpotency of $Q$ requires that
\begin{equation}
	[D_A, \partial] = 0 \quad \text{and} \quad \partial^2 = 0.
\end{equation}
This extension of the de Rham differential changes the closure \eqref{eq:44_closure} to
\begin{align}
\label{eq:ath_closure}
	0 & = s D_s \omega_\janksubfp + \sb \bar{D}_\sb \omega_\janksubfp \notag\\
			& \quad + (-1)^{s + \sb} (p + 1 - s - \sb) \partial_\psi \omega_\janksubfp \notag\\
			& \quad + 2i (-1)^{s + \sb} s \sb \omega_\janksubfs + (-1)^{p + 1} \partial \pi_\janksubfp.
\end{align}
Here $\omega$ is a spacetime $p$-form and $\pi$ a spacetime $(p + 1)$-form (both field-strengths). The first thing to notice in \eqref{eq:ath_closure} is a possible issue with applying the methods of lecture \ref{sec:lec2}. In particular, the $\partial \pi$ term has the same dimension as the torsion term and so it will be necessary for certain closure conditions on higher-degree forms to be satisfied before we can apply lemma \hyperlink{lemma1}{1} to the current form under consideration. Additionally, in order to retain the superfield structure of the $4|4$ complex we will use the exact same superform components---this is the reason for the peculiar normalizations seen in lecture \ref{sec:lec2}. However, this will require the constraints to change non-trivially. These new constraints will be those of the abelian tensor hierarchy (ATH).

Let us begin again with $p = 1$. We presume that we will eventually solve the simultaneous closure conditions $Q {\bm \omega} = 0$ for all values of $p$ and so the usual cohomological analysis carries through without issue.\footnote{It may be useful to write out some of the relevant $Q^2 = 0$ identities to see how this works. The $(QA)_{\psi \psi} = 0$ identity, for example, holds identically only after $(QF)_{s \sb \psi} = 0$ has been solved. This is generally true for all $s \leq 1$ identities but shows that lemma \hyperlink{lemma1}{1} still holds if we agree to solve the closure conditions on every part of ${\bm \omega}$ at once. Notice that this is not an issue when it comes to the obstruction complex because the obstructed forms will have a different bi-degree and are not the same forms as those we are demanding be closed.} The components of $A$ remain the same as in $4|4$ and the constraint on $U$ will therefore appear in
\begin{equation}
	0 = (QA)_{\sb \psi} = \bar{D}_\sb A_\psi - \partial_\psi A_\sb + \partial F_{\sb \psi}.
\end{equation}
Using the $4|4$ components and pulling out the $\sigma_\psi(\cdot, \sb)$ cocycle yields
\begin{align}
	0 & = - \tfrac{i}{4} \bar{D}_\sb \bar{\sigma}_\psi^{\alpha \ad} [\bar{D}_\ad, D_\alpha] U - \partial_\psi \bar{D}_\sb U - i \sb^\ad (\sigma_\psi)_{\alpha \ad} \partial W^\alpha \notag\\
		& = i \sigma_\psi \left( - \tfrac{1}{4} \bar{D}^2 D U - \partial W, \sb \right)\!.
\end{align}
Just like that, we have the first ATH constraint. Next we look at the 2-form. Because $H_{ss \psi} = H_{\sb \sb \psi} = 0$ the superfield $W_\alpha$ is still chiral.\footnote{Interestingly, we will see that this condition actually gets obstructed in the NATH.} The remaining constraint sits inside
\begin{equation}
	0 = (QF)_{s \sb \psi} = D_s F_{\sb \psi} + \bar{D}_\sb F_{s \psi} + 2i F_{\sigma(s, \sb) \psi} - \partial H_{s \sb \psi}.
\end{equation}
Pulling out the part proportional to the cocycle $\sigma_\psi(s, \sb)$ gives
\begin{align}
	0 & = - i D_s \sigma_\psi(W, \sb) - i \bar{D}_\sb \sigma_\psi(s, \bar{W}) + 2i F_{\sigma(s, \sb) \psi} + \tfrac{1}{2} \sigma_\psi(s, \sb) \partial H \notag\\
		& = \tfrac{i}{2} \sigma_\psi(s, \sb) (D^\alpha W_\alpha - \bar{D}_\ad \bar{W}^\ad - i \partial H).
\end{align}
Continuing on to the 3-form we recall that the constraint is found in
\begin{equation}
	0 = (QH)_{\sb \sb \psi \psi} = 2 \bar{D}_\sb H_{\sb \psi \psi} + \partial G_{\sb \sb \psi \psi}.
\end{equation}
The relevant cocycle here is $\bar{\sigma}_{\psi \psi}(\sb, \sb)$ and so we find
\begin{align}
	0 & = - i \bar{D}_\sb \bar{\sigma}_{\psi \psi}(\sb, \bar{D}) H - 2i \bar{\sigma}_{\psi \psi}(\sb, \sb) \partial G \notag\\
		& = - \tfrac{i}{2} \bar{\sigma}_{\psi \psi}(\sb, \sb) (\bar{D}^2 H + 4 \partial G).
\end{align}
Finally, since $G$ sits at the top of the complex its constraint remains unchanged. Thus, we have the superfield constraints
\begin{subequations}
\label{eq:ath_cons}
\begin{align}
\label{eq:ath_u_cons}
	- \tfrac{1}{4} \bar{D}^2 D_\alpha U & = \partial W_\alpha, \\
\label{eq:ath_w_cons}
	D^\alpha W_\alpha - \bar{D}_\ad \bar{W}^\ad & = i \partial H, \\
\label{eq:ath_h_cons}
	- \tfrac{1}{4} \bar{D}^2 H & = \partial G, \\
\label{eq:ath_g_cons}
	\bar{D}_\ad G & = 0.
\end{align}
\end{subequations}
Up to a different set of normalizations, these are exactly the superfield constraints in \cite{Becker:2016xgv}. The derivation of the prepotential solutions and their gauge transformations, along with the construction of the 0-form field-strength, can now be carried out as an exercise.

\hypertarget{ex4}{}
\subsection{Exercise \#4}

Use the nilpotency of $Q$ to write down the prepotentials of the ATH without carrying out a single calculation. What is the one part that must be solved ``by hand"? Is it a new calculation? Then use the prepotential solution for $U$ to construct the 0-form field-strength $E$. Finally, write down the gauge transformations for the prepotentials. (\hyperlink{sol4}{\textbf{Solution \#4}})

\vspace{10pt}
\el
\clearpage

\section{Extending the ATH to the NATH}
\label{sec:lec5}

The differential $Q$ of lecture \ref{sec:lec4} is extended further by contraction with a ``gauge" 2-form field-strength $\F$, an operation familiar from the NATH in \cite{Becker:2016rku}. The consistency relations of the NATH are shown to come from the closure of forms under this differential, with one constraint left as an exercise, and the prepotential solutions are again derived from nilpotency.

\subsection{Superfield Constraints}

The extension of the ATH to the full NATH follows much of the same logic as the extension from the $4|4$ complex to the ATH. Additionally, to keep the scope of these notes from expanding beyond our original intent we will confine ourselves to purely cohomological considerations here. In the NATH we introduce a ``gauge" 2-form field-strength $\F = \d \A - \tfrac{1}{2} [\A, \A]$ that is distinct from the ``matter" 2-form field-strength $F$. The contraction operator $\iota_\F$ with bi-degree $(2, -1)$ is then defined by taking the wedge product with $\F$ while contracting one of the internal indices. This leads to the differential $\Q$ that acts on spacetime $p$-forms as
\begin{equation}
	\Q \omega_p \ceq \D \omega_p + (-1)^{p + 1} \partial \omega_{p + 1} - (-1)^{p - 1} \iota_\F \omega_{p - 1}.
\end{equation}
This squares to zero when
\begin{equation}
\label{eq:dN_nil}
\begin{gathered}
	\D^2 = \{\partial, \iota_\F\} = \L_\F, \\
	[\D, \partial] = [\D, \iota_\F] = \partial^2 = \iota_\F^2 = 0.
\end{gathered}
\end{equation}
The super-covariant derivatives $\D$ have been gauged by the Lie derivative $\L$ with respect to the connection $\A$. The 2-form $\F$ is closed under the gauged de Rham operator and holds a chiral superfield $\W$ satisfying $\D \W = \bar{\D} \bar{\W}$. We also know that $\F_{ss} = 0$ and so the general closure condition $(\Q \omega)_\janksubfp = 0$ is
\begin{align}
	0 & = s \D_s \omega_\janksubfp + \sb \bar{\D}_\sb \omega_\janksubfp \notag\\
		& \quad + (-1)^{s + \sb} (p + 1 - s - \sb) \D_\psi \omega_\janksubfp + 2i (-1)^{s + \sb} s \sb \omega_\janksubfs \notag\\
		& \quad + (-1)^{p + 1} \partial \pi_\janksubfp + (-1)^{p - 1 + s + \sb} s (p + 1 - s - \sb) \iota_{\F_{s \psi}} \lambda_\janksubfp \notag\\
		& \quad + (-1)^{p - 1 + s + \sb} \sb (p + 1 - s - \sb) \iota_{\F_{\sb \psi}} \lambda_\janksubfp + \ldots,
\end{align}
for $\deg(\lambda, \omega, \pi) = (p - 1, p, p + 1)$. The ellipsis denotes the $\iota_{\F_{\psi \psi}}$ term that will never appear at the level of non-trivial cohomology. With this, the approach is extremely similar to that of the ATH in lecture \ref{sec:lec4}. We will use the standard de Rham components of the superforms and look at how the superfield constraints are modified by the extra terms in our new differential.

Starting with $p = 1$, the cohomology is non-trivial at the $s = 1$ level and so we look at the closure condition\footnote{Here we make use of the 0-form field-strength $E$ from \eqref{eq:0form_fs}.}
\begin{align}
	0 = (\Q A)_{\sb \psi} & = \bar{\D}_\sb A_\psi - \D_\psi A_\sb + \partial F_{\sb \psi} - \iota_{\mathcal{F}_{\sb \psi}} E \notag\\
		& = - \tfrac{i}{4} \bar{\D}_\sb \sigma_\psi^{\alpha \ad} [\bar{\D}_\ad, \D_\alpha] U + \D_\psi \bar{\D}_\sb U - i \sigma_\psi(\partial W, \sb) - \sigma_\psi(\iota_\W \partial \hat{\Phi}, \sb) \notag\\
		& = - \tfrac{i}{4} \sb_\ad \sigma_\psi^{\alpha \ad} \bar{\D}^2 \D_\alpha U - [\bar{\D}_\sb, \D_\psi] U - i \sigma_\psi(\partial W, \sb) - \sigma_\psi(\iota_\W \partial \hat{\Phi}, \sb) \notag\\
		& = \sigma_\psi \left( - \tfrac{i}{4} \bar{\D}^2 \D U + i \L_\W U - i \partial W - \iota_\W \partial \hat{\Phi}, \sb \right),
\end{align}
having used the new commutation relation
\begin{equation}
\label{eq:nath_comm_sv}
	[\bar{\D}_\ad, \D_a] = \L_{\F_{\ad a}} = - i (\sigma_a)_{\alpha \ad} \L_{\W^\alpha}.
\end{equation}
Now if we use the fact that\footnote{This is presuming that the prepotential solution for $U$ is unchanged in the NATH. This is of course true because $\iota_\F$ has bi-degree $(2, -1)$ while $U$ has $(1, 3)$.}
\begin{equation}
	i \L_{\W^\alpha} U = i \{\partial, \iota_{\W^\alpha}\} U = i \partial (\iota_{\W^\alpha} U) + \tfrac{i}{2} \iota_{\W^\alpha} \partial (\Phi + \bar{\Phi})
\end{equation}
we arrive at the first NATH constraint,
\begin{equation}
\label{eq:nath_1form_cons}
	- \tfrac{1}{4} \bar{\D}^2 \D_\alpha U = \partial (W_\alpha - \iota_{\W_\alpha} U) - \iota_{\W_\alpha} \partial \Phi.
\end{equation}
Interestingly, the terms here are not all individually chiral. This suggests that the spinor superfield of the matter 2-form is no longer chiral. And indeed,
\begin{align}
	(\Q F)_{\sb \sb \psi} & = 2 \bar{\D}_\sb F_{\sb \psi} - 2 \iota_{\F_{\sb \psi}} A_\sb \notag\\
		& = - 2i \sigma_\psi(\bar{\D}_\sb W + \iota_\W \bar{\D}_\sb U, \sb),
\end{align}
which states that the chirality of $W^\alpha$ is obstructed by
\begin{equation}
\label{eq:nath_obs_chiral}
	\bar{\D}_\ad W_\alpha = - \iota_{\W_\alpha} \bar{\D}_\ad U.
\end{equation}
However, since $\W^\alpha$ is chiral we can define the field-strength superfield
\begin{equation}
	\mathbf{W}_\alpha \ceq W_\alpha - \iota_{\W_\alpha} U
\end{equation}
which is chiral by \eqref{eq:nath_obs_chiral}. Then \eqref{eq:nath_1form_cons} becomes
\begin{equation}
	- \tfrac{1}{4} \bar{\D}^2 \D_\alpha U = \partial \mathbf{W}_\alpha - \iota_{\W_\alpha} \partial \Phi.
\end{equation}
There is additional cohomology at the $(s \sb \psi)$ level wherein we find
\begin{align}
	(\Q F)_{s \sb \psi} & = \D_s F_{\sb \psi} + \bar{\D}_\sb F_{s \psi} + 2i F_{\sigma(s, \sb) \psi} - \partial H_{s \sb \psi} - \iota_{\F_{s \psi}} A_\sb - \iota_{\F_{\sb \psi}} A_s \notag\\
		& = \tfrac{i}{2} \sigma_\psi(s, \sb) (\D^\alpha W_\alpha - \bar{\D}_\ad \bar{W}^\ad - i \partial H + \iota_{\W^\alpha} \D_\alpha U + \iota_{\bar{\W}_\ad} \bar{\D}^\ad U).
\end{align}
Now since
\begin{equation}
	\D^\alpha W_\alpha = \D^\alpha \mathbf{W}_\alpha + (\D^\alpha \iota_{\W_\alpha}) U + \iota_{\W^\alpha} \D_\alpha U
\end{equation}
and
\begin{equation}
	\bar{\D}_\ad \bar{W}^\ad = \bar{\D}_\ad \bar{\mathbf{W}}^\ad - (\bar{\D}_\ad \iota_{\bar{\W}^\ad}) U - \iota_{\bar{\W}_\ad} \bar{\D}^\ad U
\end{equation}
the constraint at the $(s \sb \psi)$ level can also be expressed as
\begin{equation}
	\D^\alpha \mathbf{W}_\alpha - \bar{\D}_\ad \bar{\mathbf{W}}^\ad = i \partial H - 2 \Omega(U \stc \W).
\end{equation}
Here we have introduced the ``Chern-Simons" superfield
\begin{equation}
	\Omega(\phi, \psi) \ceq (\D^\alpha \phi) \psi_\alpha + (\bar{\D}_\ad \phi) \bar{\psi}^\ad + \tfrac{1}{2} \phi (\D^\alpha \psi_\alpha + \bar{\D}_\ad \bar{\psi}^\ad)
\end{equation}
for $\phi$ real and $\psi$ chiral, and $\Omega(\phi \stc \psi)$ is the $*$-extended version of $\Omega$ in which the forms are multiplied with the $*$-product (contraction as $\F * \omega = \iota_\F \omega$) in addition to the usual wedge product. The constraint on the matter 3-form is left as an exercise at the end of the lecture. Finally, for the matter 4-form the cohomology is non-trivial at the $(\sb \sb \sb \psi \psi)$ level where we find
\begin{equation}
	(\Q G)_{\sb \sb \sb \psi \psi} = 3 \bar{\D}_\sb G_{\sb \sb \psi \psi} = - 6i \bar{\sigma}_{\psi \psi}(\sb, \sb) \bar{\D}_\sb G.
\end{equation}
This gives the final NATH constraint, the chirality of $G$. All together, we have the set of constraints
\begin{subequations}
\begin{align}
\label{eq:nath_U_cons}
	- \tfrac{1}{4} \bar{\D}^2 \D_\alpha U & = \partial \mathbf{W}_\alpha - \iota_{\W_\alpha} \partial \Phi, \\
\label{eq:nath_bfW_cons}
	\D^\alpha \mathbf{W}_\alpha - \bar{\D}_\ad \bar{\mathbf{W}}^\ad & = i \partial H - 2 \Omega(U \stc \W), \\
\label{eq:nath_H_cons}
	- \tfrac{1}{4} \bar{\D}^2 H & = \partial G + 2i \iota_{\W^\alpha} \mathbf{W}_\alpha, \\
	\bar{\D}_\ad G & = 0
\end{align}
\end{subequations}
for which we now wish to find a set of prepotential solutions.

\subsection{Prepotential Solutions}

Again we exploit the nilpotency of our differential to find solutions to the constraints imposed by closure. Doing so straightforwardly yields\footnote{It is worth noting how easy these solutions are to derive and how difficult they are to fully verify, especially the middle two. A similar thing occurs in higher-dimensional super-de Rham complexes where $\d^2 = 0$ gives highly non-trivial $D$-identities for free.}
\begin{subequations}
\begin{align}
\label{eq:nath_U_pre}
	U & = \tfrac{1}{2} (\Phi + \bar{\Phi}) + \partial V, \\
\label{eq:nath_bfW_pre}
	\mathbf{W}_\alpha & = - \tfrac{1}{4} \bar{\D}^2 \D_\alpha V + i \partial \Sigma_\alpha - \iota_{\W_\alpha} \Phi, \\
\label{eq:nath_H_pre}
	H & = \D^\alpha \Sigma_\alpha - \bar{\D}_\ad \bar{\Sigma}^\ad + \partial X + 2 \Omega(V \stc \W), \\
	G & = - \tfrac{1}{4} \bar{\D}^2 X + 2 \iota_{\W^\alpha} \Sigma_\alpha,
\end{align}
\end{subequations}
with $\Phi$ chiral, $V$ real, $\Sigma$ chiral, and $X$ real. For completeness, we also have
\begin{equation}
	W_\alpha = - \tfrac{1}{4} \bar{\D}^2 \D_\alpha V + i \partial \Sigma_\alpha + \iota_{\W_\alpha} (\partial V - i \hat{\Phi})
\end{equation}
as the prepotential solution for the actual superfield inside the matter 2-form. This completes the proof that the defining constraints of the non-abelian tensor hierarchy are nothing more than statements of $\Q$-closure and the solutions are found from $\Q^2 = 0$ with little effort.

\hypertarget{ex5}{}
\subsection{Exercise \#5}

Carry out the omitted calculation for the third NATH constraint. Show that the constraint can be written entirely in terms of $\mathbf{W}_\alpha$. (\hyperlink{sol5}{\textbf{Solution \#5}})

\vspace{10pt}
\el
\clearpage

\section{Composite Cohomology and CS Actions}
\label{sec:lec6}

Questions of composite cohomology are studied in the (N)ATH and cubic Chern-Simons-type actions are constructed by leveraging an isomorphism between the cohomologies of linear and bilinear superforms under the coboundary operator. One of the composite field-strengths in the NATH is left to be calculated in an exercise and additional questions concerning higher-order composites are raised.

\subsection{Cubic ACS Action}

One of the goals in \cite{Becker:2016xgv, Becker:2016rku} is to build Chern-Simons actions in $(4 + 7)|4$ that correspond to the dimensional reduction of the
\begin{equation}
	S_\text{11D} = \int C \wedge \d C \wedge \d C
\end{equation}
part of the eleven-dimensional supergravity action. These actions will be expressed in terms of the field-strength superfields and prepotential solutions that make up the (N)ATH. For simplicity, let us begin in this section with the cubic (in fields) Chern-Simons action of the ATH.

Take as a starting point the superspace Chern-Simons action
\begin{equation}
	S_\text{ACS} = \int \d^4 \theta (X u + a_0 V h) + \int \d^2 \theta (a_1 \Sigma^\alpha w_\alpha + a_2 \Phi g) + \text{h.c.},
\end{equation}
where we have suppressed the spacetime and internal measures, and recall the gauge transformations of the prepotentials are given in \eqref{eq:ath_gauge_trans}. Assume that $(u, w_\alpha, h, g)$ are gauge-invariant. It is a good check to see that the action is gauge-invariant when the coefficients are fixed so that
\begin{equation}
\label{eq:acs_action}
	S_\text{ACS} = \int \d^4 \theta (X u - V h) + \int \d^2 \theta (\Sigma^\alpha w_\alpha + \tfrac{1}{2} \Phi g) + \text{h.c.},
\end{equation}
again provided that the lowercase fields are field-strengths. However, we have not yet said what the lowercase fields actually are, only that they satisfy the closure constraints \eqref{eq:ath_cons}. This is what ensures the gauge-invariance of the action. Obviously then,
\begin{equation}
	(u, w_\alpha, h, g) = (U, W_\alpha, H, G)
\end{equation}
is a solution and generates the quadratic action. One very interesting question is then: Do there exist higher-order solutions to \eqref{eq:ath_cons} that would allow us to generate higher-order gauge-invariant Chern-Simons actions? Delightfully, this question can be answered by a cohomology analysis. Consider, for example, the 4-form $G \in \Omega^4(\mathbf{R}^{4|4} \times M)$. The simplest non-trivial closure condition is at the $(\sb \sb \sb \psi \psi)$ level and states that
\begin{equation}
\label{eq:acs_QG_sbsbsbvv}
	0 = (Q G)_{\sb \sb \sb \psi \psi} = -6i \bar{\sigma}_{\psi \psi}(\sb, \sb) \bar{D}_\sb G,
\end{equation}
requiring that $G$ be chiral. Suppose we now define the composites
\begin{subequations}
\begin{align}
	\omega_4 & \ceq E \wedge G + A \wedge H + F \wedge F, \\
	\omega_5 & \ceq A \wedge G + F \wedge H,
\end{align}
\end{subequations}
where $(E, A, F, H, G)$ are already $Q$-closed. Then
\begin{equation}
	0 \equiv (Q \omega_4)_{\sb \sb \sb \psi \psi} \sim D_\sb (\omega_4)_{\sb \sb \psi \psi} + \partial_\psi (\omega_4)_{\sb \sb \sb \psi} + \partial (\omega_5)_{\sb \sb \sb \psi \psi}.
\end{equation}
If this can be recast in the form of \eqref{eq:acs_QG_sbsbsbvv} then we are \textit{guaranteed} to find a bilinear chiral scalar $g$ hidden inside this closure condition because the non-trivial cohomology is isomorphic to the linear case involving $G$. That is, we must be able to massage the composite closure condition
\begin{align}
\label{eq:ex6_closure}
	0 \equiv (Q \omega_4)_{\sb \sb \sb \psi \psi} & = 3 \bar{D}_\sb (E G_{\sb \sb \psi \psi} - 2 F_{\sb \psi} F_{\sb \psi}- 2 A_\sb H_{\sb \psi \psi}) \notag\\
		& \quad - 3 \partial (A_\sb G_{\sb \sb \psi \psi}),
\end{align}
to look like \eqref{eq:acs_QG_sbsbsbvv} and then let $g$ be the scalar sitting in the place of $G$. The first step should then be to put the $\partial (AG)$ term into $\bar{D}_\sb$-exact form. Indeed, because of the chirality of $G$, we find that
\begin{equation}
	- 3 \partial (A_\sb G_{\sb \sb \psi \psi}) = - 6i \bar{\sigma}_{\psi \psi}(\sb, \sb) \bar{D}_\sb \partial (UG).
\end{equation}
This means that the linear and bilinear cohomologies are isomorphic at this level, and so we can find a bilinear solution $g$ to the linear superfield constraints on $G$. To find $g$, we now look under the original $\bar{D}$ to find
\begin{equation}
\label{eq:comp_g_first}
	3 E G_{\sb \sb \psi \psi} = 6 \bar{\sigma}_{\psi \psi}(\sb, \sb) \partial \hat{\Phi} G = - 3i \bar{\sigma}_{\psi \psi}(\sb, \sb) \partial (\Phi - \bar{\Phi}) G
\end{equation}
for the first term,
\begin{equation}
	- 6 F_{\sb \psi} F_{\sb \psi} = - 6 \bar{\sigma}_{\psi \psi}(\sb, \sb) W^\alpha W_\alpha
\end{equation}
for the second term, and
\begin{align}
\label{eq:lec6_acs_pr}
	- 6 A_\sb H_{\sb \psi \psi} & = - 3i \bar{D}_\sb (U \bar{\sigma}_{\psi \psi}(\sb, \bar{D}) H) + \tfrac{3i}{2} U \bar{\sigma}_{\psi \psi}(\sb, \sb) \bar{D}^2 H \notag\\
		& = \bar{\sigma}_{\psi \psi}(\sb, \sb) [- \tfrac{3i}{2} \bar{D}^2 (UH) - 6i U (\partial G)] \notag\\
		& \quad + 3i \bar{D}_\sb [(\bar{\sigma}_{\psi \psi}(\sb, \bar{D}) U) H]
\end{align}
for the third, having used \eqref{eq:ath_h_cons}. The last term in \eqref{eq:lec6_acs_pr} looks troublesome because it is not proportional to the relevant cocycle. However, notice that by being $\bar{D}_\sb$-exact it vanishes by itself inside the closure condition and does not obstruct the chirality of $g$. Now, combining all of the above terms we see that the $U \partial G$ pieces cancel and we are left with
\begin{equation}
	(Q \omega_4)_{\sb \sb \sb \psi \psi} = \bar{\sigma}_{\psi \psi}(\sb, \sb) \bar{D}_\sb (- 6i \partial \Phi G - 6 W^\alpha W_\alpha - \tfrac{3i}{2} \bar{D}^2 (UH)),
\end{equation}
where we have used \eqref{eq:ath_u_sol}. Comparing this to  \eqref{eq:acs_QG_sbsbsbvv} we find
\begin{equation}
\label{eq:acs_g_us}
	g = - i W^\alpha W_\alpha + \tfrac{1}{4} \bar{D}^2 (UH) + \partial \Phi G
\end{equation}
for the 4-form bilinear composite.

A similar procedure can be applied to get the remaining composites as well. For the 1-form $u$, we end up with a $\bar{D}_\sb$-exact expression that includes two terms, $i \partial \hat{\Phi} U$ and $- \tfrac{1}{2} U \partial U$. The second has the incorrect bi-degree so only the first enters $u$. The analysis for the 2-form $w_\alpha$ is very similar to that of $g$, having to throw away two terms that vanish separately from the chirality constraint on $w_\alpha$. The 3-form $h$ is the most difficult, due to the reality of the composite (and the fact that the relevant linear closure condition is not simply a $\bar{D}_\sb$-exact expression but also has a number of $\im \delta$ terms). Since reality is not a cohomological condition but something imposed by hand for the sake of irreducibility, we have to be sure to extract only the real part of the composite. Actually, since $u$ is imaginary, we must pull out the imaginary part of the 3-form composite closure condition and assign that to $h$. Then re-scaling all the composites by an aesthetic factor of $2i$, we arrive at\footnote{\textbf{Note added:} After giving these lectures, it was pointed out by Daniel Robbins that this procedure for constructing composite superfields is ambiguous for higher-order cases. This is effectively because we are working with the closure conditions (identically vanishing) instead of the forms themselves. A revised procedure that uses the forms directly is given in \cite{Becker:2017njd}. The logic is very similar to what we first tried here---we still try to ``match" composite superforms to their linear counterparts---but has the advantage of being completely unambiguous and universal.}
\begin{subequations}
\label{eq:ath_comps}
\begin{align}
	u & = - 2 \partial \hat{\Phi} U, \\
	w_\alpha & = \tfrac{i}{2} \bar{D}^2 (U D_\alpha U) + 2i \partial \Phi W_\alpha, \\
	h & = 4 \Omega(U, W) - 2 \partial \hat{\Phi} H, \\
	g & = 2 W^\alpha W_\alpha + \tfrac{i}{2} \bar{D}^2 (U H) + 2i \partial \Phi G.
\end{align}
\end{subequations}
With these, the action \eqref{eq:acs_action} is gauge-invariant.

\subsection{Additional Applications}

This procedure is very naturally extended to the full NATH. In that case we do need to correct for the non-chiral 2-form field-strength, but it can be shown with slightly more effort that we recover the composites \eqref{eq:ath_comps} with the chiral $\mathbf{W}_\alpha$ and an additional term proportional to $(\iota_{\W^\alpha} U) \D_\alpha U + \text{h.c.}$ inside the composite 3-form $h$. Such a result is consistent with the action given in equation (6.13) of \cite{Becker:2016rku}.

\hypertarget{ex6}{}
\subsection{Exercise \#6}

Calculate $g$ for the cubic NACS action by modifying the corresponding derivation for the cubic ACS action. Use the composite closure condition
\begin{align}
\label{eq:ex6_closure}
	0 \equiv (\Q \omega_4)_{\sb \sb \sb \psi \psi} & = 3 \bar{\D}_\sb (E G_{\sb \sb \psi \psi} - 2 F_{\sb \psi} F_{\sb \psi} - 2 A_\sb H_{\sb \psi \psi}) \notag\\
		& \quad - 3 \partial (A_\sb G_{\sb \sb \psi \psi}) + 12 \iota_{\F_{\sb \psi}} (A_\sb F_{\sb \psi}),
\end{align}
which is easily verified from the linear closure conditions. Be sure to check that the necessary isomorphism exists by ensuring that the composite closure condition can be expressed in $\bar{\D}_\sb$-exact form. It will be useful later in the calculation if you begin expanding $W_\alpha$ into the chiral and non-chiral pieces,
\begin{equation}
\label{eq:ex6_W_exp}
	W_\alpha = \mathbf{W}_\alpha + \iota_{\W_\alpha} U,
\end{equation}
as soon as possible. Once the isomorphism is established, pull out the appropriate cocycle and identify the composite $g$. If done correctly, you should have a leftover term that vanishes by itself in the closure condition (just as in the abelian case). Make sure that the result is entirely in terms of the chiral $\mathbf{W}_\alpha$, as claimed in the lecture. (\hyperlink{sol6}{\textbf{Solution \#6}})

\vspace{10pt}
\el
\clearpage

\renewcommand*{\refname}{\vspace*{-1em}}
\section*{Final Comments}
\addcontentsline{toc}{section}{Final Comments}
\vspace*{-1.1em}
\label{sec:final}
\vspace{8pt}

Over the course of these lectures we have hopefully managed to present a more coherent and computationally useful perspective on superspace cohomology than is usually found in textbooks. Doing so allowed us to reduce the (N)ATH constructions in \cite{Becker:2016xgv, Becker:2016rku} to little more than simple cohomological calculations. While this was certainly a very nice interpretation and reproduction of prior results, it would be of little interest if we could not leverage it to learn something new. The tensor hierarchies laid out here were motivated by the reduction of the 3-form gauge field in eleven-dimensional supergravity and were all built in flat superspaces. However, reducing the graviton and gravitino to four dimensions as well will require that we curve those superspaces. One relevant question is then: How do we couple the NATH to gravity?

From a cohomological point of view, this is now an extremely simple question to answer. As explained in \cite{Linch:2014iza, Randall:2014gza} the additional (dimensionful) torsions that arise in curved superspaces do not affect the constraint analysis based on $\delta$. This ensures that the supergravity fields enter the constraints in a very controlled and easily understandable way. For example, if you wish to couple the NATH to the so-called ``old-minimal" supergravity (the formulation laid out in chapter XV of WB) to the NATH, you will find that the only thing that changes at the linear level (which consists of the field-strength constraints, prepotential solutions, and gauge variations for $U$, $W_\alpha$ (or $\mathbf{W}_\alpha$), $H$, and $G$) is that $\bar{\D}^2 \mapsto \mathscr{\bar{D}}^2 - 8R$, consistent with the recent construction in \cite{Aoki:2016rfz}. Here $\mathscr{D}$ is the fully covariant (diffeo + gauge + super) derivative and $R$ is the chiral superfield holding the Ricci scalar as a component field. In the future, we would like to use this insight to simplify the construction of gauge-invariant Chern-Simons actions in curved 4D, $N = 1$ superspace.

We now have a significantly improved toolset for studying completely general tensor hierarchies. The linear level is totally trivialized. Associated composite field-strengths can be found with more effort (through the method of \cite{Becker:2017njd}), although still in a manner far simpler than the approach of \cite{Becker:2016rku}. Such an improved toolset will, we hope, expedite our study of supergravity reductions and can apply more generally to any future research projects on supersymmetric gauge theories in arbitrary superspaces.

Finally, please feel free to \href{mailto:srandall@berkeley.edu}{contact me} with questions, comments, or corrections concerning these lecture notes.

\vspace{10pt}
\el
\clearpage

\appendix
\numberwithin{equation}{subsection}
\section{Exercise Solutions}
\label{sec:sols}

Collected solutions to the lecture exercises.

\hypertarget{sol1}{}
\subsection{Solution \#1}

(\hyperlink{ex1}{\textbf{Exercise \#1}}) This is exercise XIII.8 from WB and we make use of the hints given in their version of the problem statement. These were omitted from our version of the exercise (1) to more accurately simulate how it feels to try to come up with a way of checking one of the top closure conditions on your own and (2) because they have a sign error in their hints that would have led you to the wrong answer. We note from \eqref{eq:conv_dim2_ssbv} that
\begin{equation}
	F_{ac} = - \tfrac{i}{4} \bar{\sigma}_a^{\bd \alpha} (\bar{D}_\bd F_{\alpha c} + D_\alpha F_{\bd c}).
\end{equation}
Acting on this with a spacetime derivative yields
\begin{equation}
	\partial_b F_{ac} = - \tfrac{i}{4} \bar{\sigma}_a^{\bd \alpha} [\bar{D}_\bd (D_\alpha F_{bc} + \partial_c F_{\alpha b}) + D_\alpha (\bar{D}_\bd F_{bc} + \partial_c F_{\bd b})],
\end{equation}
where we have used the dimension-$\tfrac{5}{2}$ closure conditions\footnote{If you enjoy these checks, you could of course verify that \eqref{eq:conv_dim52_closure} follows from the already-derived components and constraints as well. I do not, and so will refrain from doing so here.}
\begin{subequations}
\label{eq:conv_dim52_closure}
\begin{align}
	0 & = D_\alpha F_{bc} + \partial_b F_{c \alpha} + \partial_c F_{\alpha b}, \\
	0 & = \bar{D}_\ad F_{bc} + \partial_b F_{c \ad} + \partial_c F_{\ad b}.
\end{align}
\end{subequations}
Remember the graded anti-symmetry of the indices on these components when plugging them in (the corresponding formula in WB has the incorrect sign). The first and third terms combine as
\begin{equation}
	- \tfrac{i}{4} \bar{\sigma}_a^{\bd \alpha} (\bar{D}_\bd D_\alpha + D_\alpha \bar{D}_\bd) F_{bc} = \partial_a F_{bc}.
\end{equation}
The second and fourth terms yield
\begin{equation}
	- \tfrac{i}{4} \bar{\sigma}_a^{\bd \alpha} \partial_c (\bar{D}_\bd F_{\alpha b} + D_\alpha F_{\bd b}) = \partial_c F_{ab},
\end{equation}
having used \eqref{eq:conv_dim2_ssbv}. Finally, exploiting the anti-symmetry of $F$ gives
\begin{equation}
	\partial_b F_{ac} + \partial_a F_{cb} + \partial_c F_{ba} = 0,
\end{equation}
and so the dimension-3 closure condition is indeed automatically satisfied.

This is a nice exercise to familiarize ourselves with basic computations in $4|4$. The true point of the problem, however, is to show that it can sometimes be difficult to know how to begin looking at a given closure condition. These are the kinds of things we wish to address in lecture \ref{sec:lec2}.
\clearpage

\hypertarget{sol2}{}
\subsection{Solution \#2}

(\hyperlink{ex2}{\textbf{Exercise \#2}}) Since $4|4$ is a principal superspace we know by definition that one of the cocycles is $\sigma_\psi(s, \bar{\xi})$ for an arbitrary barred Weyl spinor $\bar{\xi}$. We also know that the cocycles, being intrinsic to the algebra, are built entirely out of the Pauli matrices and the $j^A$. In four dimensions, we could thus have three possible combinations: the scalar, the vector, and the 2-form, with the higher-form combinations being dual to one of these. The scalar is trivial since its annihilation by $\delta$ is due simply to its lack of a vector index to contract upon. The vector cocycle is the one immediately given from principality, and so we can ask if a 2-form cocycle exists. The simplest way to construct one is to take
\begin{equation}
	(\bar{\xi}_b)_\ad = (\sigma_b s)_\ad
\end{equation}
so that
\begin{equation}
	0 = \delta \sigma_a(s, \bar{\xi}) = \delta s^\alpha (\sigma_a)_{\alpha \ad} s^\beta (\sigma_b)_{\beta \bd} \epsilon^{\ad \bd} \propto \delta \sigma_{ab}(s, s),
\end{equation}
where we have used the symmetry of the $s$-variables to enforce the anti-symmetry of the vector indices and we have introduced the notation
\begin{equation}
	\sigma_{\psi \psi}(s, s) \ceq s^\alpha (\sigma_{\psi \psi})_\alpha{}^\beta s_\beta.
\end{equation}
Thus, we can conclude that
\begin{equation}
	H_\delta \supset \{1, \sigma_\psi(s, \bar{\xi}), \sigma_{\psi \psi}(s, s)\},
\end{equation}
where 1 denotes the scalar cocycle. Additionally, the conjugate versions of these, $\sigma_\psi(\xi, \sb)$ and $\bar{\sigma}_{\psi \psi}(\sb, \sb)$, are also cocycles of the Lie algebra and so
\begin{equation}
\label{eq:44_cocycles}
	H_\delta = \{1, \sigma_\psi(s, \bar{\xi}), \sigma_\psi(\xi, \sb),\sigma_{\psi \psi}(s, s), \bar{\sigma}_{\psi \psi}(\sb, \sb) \}.
\end{equation}
Finally, note that the vector cocycle only requires a single $s$ or $\sb$ to be in the kernel of $\delta$ while the 2-form requires that both of its indices be symmetrized.

Thus far we have been fairly cavalier about the difference between $H_\delta$ and $\ker \delta$. We have tried to avoid terms that are obviously in $\im \delta$ but we have not been quite careful enough. There is no ambiguity to the scalar or 2-form cocycles, but there is a subtlety to be aware of when using the principal cocycle with $\xi = s$. In that case $\sigma_\psi(s, \sb)$ by itself is certainly a member of $H_\delta$, but
\begin{equation}
\label{eq:coh_to_im_el}
	k_s \sigma_\psi(s, \sb) \propto \sigma^a(s, \sb) \sigma_{a \psi}(s, k).
\end{equation}
Thus, when the vector cocycle is multiplied by a symmetrized spinor $k$ (barred or unbarred) it moves from $H_\delta$ to $\im \delta$. The fact that such a combination is \textit{not} an element of the cohomology is relevant to the analysis of the 3-form field-strength in $4|4$. We will point out where this happens in lecture \ref{sec:lec3}.
\clearpage

\hypertarget{sol3}{}
\subsection{Solution \#3}

(\hyperlink{ex3}{\textbf{Exercise \#3}}) Let us generate the ``alternative 3-form" $H'$ from the 2-form $F$ through $\d F = H'$ with
\begin{equation}
	D_\alpha \bar{W}_\ad - \bar{D}_\ad W_\alpha = H_{\alpha \ad}.
\end{equation}
This obstruction occurs at the $(ss \psi)$ level and so we find the lowest component
\begin{equation}
	H'_{ss \psi} = \sigma_{\psi a}(s, s) H^a,
\end{equation}
with $H'_{sss} = H'_{ss \sb} = H'_{s \sb \sb} = H'_{\sb \sb \sb} = H'_{s \sb \psi} = 0$. The dimension-3 constraint on $H_a$ must sit inside the $(ss \psi \psi)$ closure condition since the 2-form cocycle has the same type of spinor indices. If we now also use the given component \eqref{eq:hprime_svv} we find that the closure condition yields
\begin{align}
	(\d H)_{ss \psi \psi} & = 2 D_s H'_{s \psi \psi} + 2 \partial_\psi H'_{ss \psi} \notag\\
		& = - \tfrac{i}{4} D_s \sigma_\psi(s, \bar{D}) H_\psi + 2 \partial_\psi \sigma_{\psi a}(s, s) H^a \notag\\
		& = - \tfrac{i}{16} \sigma_{\psi \psi}(s, s) D_\alpha \bar{D}_\ad H^{\alpha \ad}
\end{align}
and so the dimension-3 constraint is
\begin{equation}
\label{eq:hprime_dim3_cons}
	D_\alpha \bar{D}_\ad H^{\alpha \ad} = 0 ~~\Rightarrow~~ [D_\alpha, \bar{D}_\ad] H^{\alpha \ad} = \partial_{\alpha \ad} H^{\alpha \ad} = 0.
\end{equation}
It should be emphasized how little calculation needs to be done here; all that's necessary is to pull out the anti-symmetric part of two pairs of spinor indices. In fact, if you try to check the entire identity explicitly you will see that it doesn't work; the symmetrized parts of the commutator half of $D \bar{D}$ remain. What we omitted by giving the component \eqref{eq:hprime_svv} was that there is actually a constraint on $H_a$ at the $(ss \sb \psi)$ level, unlike what happened for the original 3-form $H$. Now, it is certainly possible to go back and get that constraint and use it in the determination of \eqref{eq:hprime_dim3_cons}. However, this is exactly the kind of brute-force approach that we have been trying to avoid since the first lecture because it is this brute-forcing that will end up causing fits in more complicated superspaces. Instead, we noted that we can pull out a cocycle from the first term in the closure condition and so whatever sits behind that must be a constraint and the rest must cancel automatically. The key here is to not trouble ourselves with what eventually drops out of the closure condition and to instead focus squarely on the pieces that end up being interesting.

We are also asked about obstructing the closure of $H'$ with a 4-form. We see that we will indeed generate a scalar superfield $G$ and so the complex does join back up instead of continuing to branch. Notice that this had to happen because of the cocycle structure of the algebra. For \textit{any} 3-form in this superspace the highest-dimension constraint must come at the $(ss \psi \psi)$ level. Since the sole 2-form cocycle has exactly that index structure the constraint must end up with no free indices and obstructing that constraint would generate a scalar superfield for the 4-form. This argument extends more generally to any superspace and we conclude by the same arguments that the complex will never fully branch and can at most form bubbles. An illustration of this in five dimensions is given in figure 3 of \cite{Linch:2014iza}. Another application is in $4|8$. It is remarked in the final line of \cite{Gates:1980ay} that the two inequivalent vector multiplets might each generate a new tensor multiplet of their own. As we see here, that is not true; the structure of the constraints will be identical.

You may object to this conclusion by arguing that the closure of $H'$ produced non-trivial cohomology prior to the dimension-3 constraint. Why, then, could we not obstruct that constraint instead of \eqref{eq:hprime_dim3_cons}? In general, the motivation for obstructing the highest closure condition is to make the the lowest component of the next form be the only component proportional to a cocycle. This ensures that only a single superfield ever enters the form components and that the multiplet remains irreducible.

\hypertarget{sol4}{}
\subsection{Solution \#4}

(\hyperlink{ex4}{\textbf{Exercise \#4}}) We now want to use the nilpotency of $Q$ to find the solutions to the ATH constraints \eqref{eq:ath_cons}. Starting at the top, $Q^2 = 0$ implies that $G = - \tfrac{1}{4} \bar{D}^2 X$ where $H = \partial X + K$ for $K \in \ker \bar{D}^2$. Using $Q^2 = 0$ at one degree lower fixes $K = D \Sigma - \bar{D} \bar{\Sigma}$ where $\Sigma$ is chiral and $W_\alpha = i \partial \Sigma_\alpha + L_\alpha$ where $L$ satisfies $DL = \bar{D} \bar{L}$. Continuing downwards, another application of $Q^2 = 0$ fixes $L$ to be $L_\alpha = - \tfrac{1}{4} \bar{D}^2 D_\alpha V$ for a real scalar $V$ so that $U = \partial V + M$ where $M \in \ker \bar{D}^2 D_\alpha$. Determination of $M$ is the sole thing that must be completed without the help of $Q^2 = 0$ and it was already achieved in the second half of lecture \ref{sec:lec3} as $M = \tfrac{1}{2} (\Phi + \bar{\Phi})$ for $\Phi$ chiral. Thus, we have the solutions
\begin{subequations}
\label{eq:ath_sols}
\begin{align}
\label{eq:ath_u_sol}
	U & = \tfrac{1}{2} (\Phi + \bar{\Phi}) + \partial V, \\
\label{eq:ath_w_sol}
	W_\alpha & = - \tfrac{1}{4} \bar{D}^2 D_\alpha V + i \partial \Sigma_\alpha, \\
	H & = D^\alpha \Sigma_\alpha - \bar{D}_\ad \bar{\Sigma}^\ad + \partial X, \\
	G & = - \tfrac{1}{4} \bar{D}^2 X,
\end{align}
\end{subequations}
for $\Phi$ chiral, $V$ real, $\Sigma$ chiral, and $X$ real. This is very similar to the set of prepotentials for $4|4$ with the only difference being the new requirement that $X$ be real instead of entirely unconstrained. Also, notice that \textit{nothing} simultaneously non-trivial and new needed to be worked out. Everything came for free from $Q^2 = 0$ except for the solution to $\bar{D}^2 D_\alpha U = 0$, but that we already got from the standard $4|4$ analysis. This occurred because the complex $\Omega^\bullet(\mathbf{R}^{4|4} \times M)$ is merely an extension of the super-de Rham complex and not something entirely new. We will use this line of thinking again when we construct the prepotentials for the NATH in lecture \ref{sec:lec5}. Take note of the fact that we are also not checking these prepotentials (beyond any cursory checks to make sure the signs are okay). While these solutions are certainly not difficult to check, the fact that we have a method for avoiding the checks in general will be massively valuable when we begin looking at higher-order solutions to the superfield constraints.

To build the 0-form field-strength $E$ we use \eqref{eq:ath_u_sol} in
\begin{subequations}
\begin{align}
	0 & = (Q E)_s = D_s E - \partial A_s = D_s (E - \tfrac{1}{2} \partial \Phi), \\
	0 & = (Q E)_\sb = \bar{D}_\sb E - \partial A_\sb = \bar{D}_\sb (E + \tfrac{1}{2} \partial \bar{\Phi}),
\end{align}
\end{subequations}
so that
\begin{equation}
\label{eq:0form_fs}
	E = \tfrac{1}{2} \partial (\Phi - \bar{\Phi}) = i \partial \hat{\Phi},
\end{equation}
where
\begin{equation}
	\hat{\Phi} \ceq \tfrac{1}{2i} (\Phi - \bar{\Phi}).
\end{equation}
This superform will be tremendously useful later when we begin building composite superforms in lecture \ref{sec:lec6}.

Finally, the gauge variations of the prepotentials can be obtained from a second round of $Q^2 = 0$ applications. Following the $4|4$ analysis, we find
\begin{subequations}
\label{eq:ath_gauge_trans}
\begin{align}
	\delta X & = D^\alpha \lambda_\alpha - \bar{D}_\ad \bar{\lambda}^\ad, \\
	\delta \Sigma_\alpha & = - \tfrac{i}{4} \bar{D}^2 D_\alpha L - \partial \lambda_\alpha, \\
	\delta V & = - \tfrac{1}{2} (\Lambda + \bar{\Lambda}) + \partial L, \\
	\delta \Phi & = \partial \Lambda,
\end{align}
\end{subequations}
as the prepotential transformations in the ATH. It is also worth pointing out that taking $\partial \rightarrow 0$ reduces $Q \rightarrow \d$ and all our results reduce to those of the $4|4$ de Rham complex. For this reason we will occasionally refer to the $4|4$ complex as the \textit{truncated ATH}.

\hypertarget{sol5}{}
\subsection{Solution \#5}

(\hyperlink{ex5}{\textbf{Exercise \#5}}) For the matter 3-form, the cohomology is non-trivial at the $(\sb \sb \psi \psi)$ level. Here we have
\begin{align}
	(\Q H)_{\sb \sb \psi \psi} & = 2 \bar{\D}_\sb H_{\sb \psi \psi} + \partial G_{\sb \sb \psi \psi} + 4 \iota_{\F_{\sb \psi}} F_{\sb \psi} \notag\\
		& = - i \bar{\D}_\sb \bar{\sigma}_{\psi \psi}(\sb, \bar{\D}) H - 2i \bar{\sigma}_{\psi \psi}(\sb, \sb) \partial G - 4 \sigma_\psi(\iota_\W, \sb) \sigma_\psi(W, \sb) \notag\\
		& = - \tfrac{i}{2} \bar{\sigma}_{\psi \psi}(\sb, \sb) (\bar{\D}^2 H + 4 \partial G + 8i \iota_{\W^\alpha} W_\alpha)
\end{align}
and so we discover the third NATH constraint,
\begin{equation}
	- \tfrac{1}{4} \bar{\D}^2 H = \partial G + 2i \iota_{\W^\alpha} W_\alpha = \partial G + 2i \iota_{\W^\alpha} \mathbf{W}_\alpha
\end{equation}
since $\iota_\W \iota_\W = 0$. Again, we see that the constraint can be expressed in terms of the chiral combination $\mathbf{W}_\alpha$, consistent with \cite{Becker:2016rku}.

\hypertarget{sol6}{}
\subsection{Solution \#6}

(\hyperlink{ex6}{\textbf{Exercise \#6}}) We begin by looking at the non-$\bar{\D}_\sb$-exact terms in the composite closure condition \eqref{eq:ex6_closure}. Calculating we find
\begin{equation}
	- 3 \partial (A_\sb G_{\sb \sb \psi \psi}) = - 6i \bar{\sigma}_{\psi \psi}(\sb, \sb) \bar{\D}_\sb \partial (UG)
\end{equation}
and
\begin{equation}
	12 \iota_{\F_{\sb \psi}} (A_\sb F_{\sb \psi}) = 12 \bar{\sigma}_{\psi \psi}(\sb, \sb) \bar{\D}_\sb (\iota_{\W^\alpha} (U \mathbf{W}_\alpha) + \tfrac{1}{2} (\iota_{\W^\alpha} U)(\iota_{\W_\alpha} U)).
\end{equation}
In the second calculation we have used
\begin{equation}
	\sb^\ad \sb^\bd (\sigma_\psi)_{\alpha \ad} (\sigma_\psi)_{\beta \bd} = - \epsilon_{\alpha \beta} \bar{\sigma}_{\psi \psi}(\sb, \sb)
\end{equation}
as well as \eqref{eq:nath_obs_chiral} and the fact that $\iota_\W \iota_\W = 0$. Since all the terms are $\bar{\D}_\sb$-exact, the isomorphism to the linear case is established and we now need to extract the composite. Looking under the original $\bar{\D}_\sb$ we find \eqref{eq:comp_g_first} for the first term,
\begin{align}
	- 6 F_{\sb \psi} F_{\sb \psi} & = - 6 \bar{\sigma}_{\psi \psi}(\sb, \sb) W^\alpha W_\alpha \notag\\
		& = - 6 \bar{\sigma}_{\psi \psi}(\sb, \sb) [\mathbf{W}^\alpha \mathbf{W}_\alpha + \iota_{\W^\alpha} U (2 \mathbf{W}_\alpha + \iota_{\W_\alpha} U)]
\end{align}
for the second term, and
\begin{align}
\label{eq:ex6_pr}
	- 6 A_\sb H_{\sb \psi \psi} & = - 3i \bar{\D}_\sb (U \bar{\sigma}_{\psi \psi}(\sb, \bar{\D}) H) + \tfrac{3i}{2} U \bar{\sigma}_{\psi \psi}(\sb, \sb) \bar{\D}^2 H \notag\\
		& \sim \bar{\sigma}_{\psi \psi}(\sb, \sb) [- \tfrac{3i}{2} \bar{\D}^2 (UH) - 6i U (\partial G + 2i \iota_{\W^\alpha} \mathbf{W}_\alpha)]
\end{align}
for the third, having used \eqref{eq:nath_H_cons} and tossing the $\bar{\D}_\sb$-exact term
\begin{equation}
	- 6 A_\sb H_{\sb \psi \psi} = \eqref{eq:ex6_pr} + 3i \bar{\D}_\sb [(\bar{\sigma}_{\psi \psi}(\sb, \bar{\D}) U) H]
\end{equation}
that vanishes by itself in the closure condition. Now, combining all of the above terms we see that the $(\iota_\W U)^2$ terms cancel, as well as the $\iota_\W (U \mathbf{W})$ parts and the $U \partial G$ pieces. All together, we're left with
\begin{equation}
	(\Q \omega_4)_{\sb \sb \sb \psi \psi} = \bar{\sigma}_{\psi \psi}(\sb, \sb) \bar{\D}_\sb (- 6i \partial \Phi G - 6 \mathbf{W}^\alpha \mathbf{W}_\alpha - \tfrac{3i}{2} \bar{\D}^2 (UH)),
\end{equation}
where we have used \eqref{eq:nath_U_pre}. Comparing this to the non-abelian version of \eqref{eq:acs_QG_sbsbsbvv} we find
\begin{equation}
	g = \partial \Phi G - i \mathbf{W}^\alpha \mathbf{W}_\alpha + \tfrac{1}{4} \bar{\D}^2 (UH).
\end{equation}
This is the non-abelian version of \eqref{eq:acs_g_us}, and requires no structural changes beyond $D \rightarrow \D$ and $W \rightarrow \mathbf{W}$.

\vspace{10pt}
\el
\clearpage

\section{Useful Superspace Identities}
\numberwithin{equation}{section}
\label{sec:ids}

In this appendix we collect a variety of useful 4D, $N = 1$ superspace relations between the covariant derivatives and the $\sigma$-matrices.

\vspace{4pt}
\textbf{Derivative identities}: Taken from \S2.5.6 of Buchbinder and Kuzenko's comprehensive text \cite{Buchbinder:1995uq}:
\begin{gather}
	D_\alpha D_\beta = \tfrac{1}{2} \epsilon_{\alpha \beta} D^\gamma D_\gamma = \tfrac{1}{2} \epsilon_{\alpha \beta} D^2 \\
	\bar{D}_\ad \bar{D}_\bd = - \tfrac{1}{2} \epsilon_{\ad \bd} \bar{D}_\gd \bar{D}^\gd = - \tfrac{1}{2} \epsilon_{\ad \bd} \bar{D}^2 \\
	D_\alpha D_\beta D_\gamma = \bar{D}_\ad \bar{D}_\bd \bar{D}_\gd = 0 \\
	\{D_\alpha, \bar{D}_\ad\} = - 2i \partial_{\alpha \ad} \\
	\{D_\alpha, D_\beta\} = \{\bar{D}_\ad, \bar{D}_\bd\} = [D_A, \partial_a] = 0 \\
	[D^2, \bar{D}_\ad] = - 4i \partial_{\alpha \ad} D^\alpha \\
	[\bar{D}^2, D_\alpha] = 4i \partial_{\alpha \ad} \bar{D}^\ad \\
	D^\alpha \bar{D}^2 D_\alpha = \bar{D}_\ad D^2 \bar{D}^\ad
\end{gather}
\textbf{Pauli matrix identities}: Taken from appendix B of WB:
\begin{gather}
	\sigma_{\alpha \ad}^a \bar{\sigma}_a^{\beta \bd} = - 2 \delta_\alpha^\beta \delta_\ad^\bd, \\
	\operatorname{Tr} \sigma^a \bar{\sigma}^b = - 2 \eta^{ab} \\
	v_{\alpha \ad} \ceq \sigma_{\alpha \ad}^a v_a \\
	v^a = - \tfrac{1}{2} (\bar{\sigma}^a)^{\alpha \ad} v_{\alpha \ad} \\
	(\sigma^{ab})_\alpha{}^\beta = \tfrac{1}{4} [\sigma_{\alpha \ad}^a (\bar{\sigma}^b)^{\ad \beta} - \sigma_{\alpha \ad}^b (\bar{\sigma}^a)^{\ad \beta}] \\
	(\bar{\sigma}^{ab})^\ad{}_\bd = \tfrac{1}{4} [(\bar{\sigma}^a)^{\alpha \ad} \sigma^b_{\alpha \bd} - (\bar{\sigma}^b)^{\alpha \ad} \sigma^a_{\alpha \bd}] \\
	(\sigma_{ab})_\alpha{}^\alpha = (\bar{\sigma}_{ab})^\ad{}_\ad = 0 \\
	\sigma^a_{\alpha \ad} (\sigma_{ab})_\beta{}^\gamma = \tfrac{1}{2} (\sigma_b)_{\ad \beta} \delta_\alpha^\gamma - \tfrac{1}{2} \epsilon_{\alpha \beta} (\sigma_b)_\ad{}^\gamma \\
	\sigma^a_{\alpha \ad} (\bar{\sigma}_{ab})^\bd{}_\gd = - \tfrac{1}{2} (\sigma_b)_{\alpha \gd} \delta_\ad^\bd + \tfrac{1}{2} \epsilon_{\ad \gd} (\sigma_b)_\alpha{}^\bd
\end{gather}

\vspace{10pt}
\el
\clearpage

\clearpage

\renewcommand*{\refname}{\vspace*{-1em}}

\end{document}